\documentclass{mn2e}

\usepackage{graphicx}
\usepackage{psfig}
\newcommand{\figstart}[1]  {\begin{figure} \psfig{#1}}
\newcommand{\lfigstart}[1] {\begin{figure*} \psfig{#1}}
\newcommand{\figend} {\end{figure}}
\newcommand{\lfigend} {\end{figure*}}

\newcommand{\dist} {$d_{21.5}$}
\newcommand{\tiger} {{\tt TIGER}}
\newcommand{\kms} {$\mbox{km s}^{-1}$}
\newcommand{\kmskpc} {$\mbox{km s}^{-1}\;\mbox{kpc}^{-1}$}
\newcommand{\Msun} {$\mbox{M}_{\sun}$}
\newcommand{\Lsun} {$\mbox{L}_{\sun}$}

\newcommand{\Ag} {\AA~} 

\newcommand{\ergcmsa}{erg~s$^{-1}$~cm$^{-2}$~arcsec$^{-2}$}
\newcommand{\ergcmsaa}{erg~s$^{-1}$~cm$^{-2}$~arcsec$^{-2}$~\AA$^{-1}$}

\newcommand{\lda}{$\lambda$}
\newcommand{\mgb}{Mg$b$}

\newcommand{\NI}{[{\sc N$\,$i}]}

\newcommand{\red}{\mbox{\tt red}}
\newcommand{\Red}{\mbox{\tt Red}}
\newcommand{\blue}{\mbox{\tt blue}}
\newcommand{\Blue}{\mbox{\tt Blue}}
\newcommand{\Fem}{$<$Fe$>$}
\newcommand{\Ha}{\hbox{H$\alpha$}}
\newcommand{\NII}{\hbox{[N\,{\sc ii}]}}

\newcommand{\NIIwb}{\hbox{[N\,{\sc ii}]$\lambda $6583}}
\newcommand{\NIIww}{\hbox{[N\,{\sc ii}]$\lambda\lambda $6548,6583}}
\newcommand{\SII}{\hbox{[S\,{\sc ii}]}}
\newcommand{\SIIwa}{\hbox{[S\,{\sc ii}]$\lambda $6717}}
\newcommand{\SIIwb}{\hbox{[S\,{\sc ii}]$\lambda $6731}}
\newcommand{\SIIww}{\hbox{[S\,{\sc ii}]$\lambda\lambda $6717,6731}}
\newcommand{\cmc}{cm$^{-3}$}
\newcommand{\ten}[1] {10$^{#1}$}
\newcommand{\aj}{AJ}         
\newcommand{\aaa}{A\&A}      
\newcommand{\apj}{ApJ}       
\newcommand{\mnras}{MNRAS}   
\newcommand{\nat}{Nat}       

\def\etal{et al.~}
\def\deg{^{\circ}}

\def\spose#1{\hbox to 0pt{#1\hss}}
\def\lta{\mathrel{\spose{\lower 3pt\hbox{$\sim$}}
    \raise 2.0pt\hbox{$<$}}}
\def\gta{\mathrel{\spose{\lower 3pt\hbox{$\sim$}}
    \raise 2.0pt\hbox{$>$}}}
\newdimen\hssize
\newdimen\hdsize
\title[An inner bar in the early-type galaxy NGC~2974]
{A two-arm gaseous spiral in the inner 200~pc of the early-type galaxy NGC 2974:
signature of an inner bar} 
\author[E. Emsellem, P. Goudfrooij, and P. Ferruit]{
Eric Emsellem,$^{1}$, 
Paul Goudfrooij,$^2$ and 
Pierre Ferruit$^1$ \\
$^1$ Centre de Recherche Astronomique de Lyon, Observatoire de Lyon, 9
 av. Charles-Andr\'e, 69561 Saint-Genis Laval Cedex, France \\
$^2$ Space Telescope Science Institute, 3700 San Martin Drive,
 Baltimore, MD 21218, USA}

\date{Accepted .
      Received}

\pagerange{\pageref{firstpage}--\pageref{lastpage}}
\pubyear{2003}

\begin{document}

\maketitle

\label{firstpage}

\begin{abstract}
\tiger\ integral-field spectrography and {\it Hubble Space Telescope\/}
WFPC2 imaging of the E3 galaxy NGC~2974 are used to derive the kinematics of
the stellar and ionized gas components in its central 500~pc.  We derive
a numerical two-integral distribution function from a Multi-Gaussian Expansion
(MGE) mass model using the Hunter \& Qian formalism.  The \tiger\ as well as
published long-slit stellar kinematics, including higher order moments, are
well fitted with this self-consistent model, requiring neither the addition of
a significant mass contribution from a hidden disc structure, nor the presence
of a central dark mass (at that spatial resolution).  The data reveal the
presence of a striking, highly contrasted, two-arm gaseous spiral structure
within a radius of $\sim$\,200~pc, corresponding to a total mass of 6.8 $\times$
\ten{4}\ solar masses of ionized gas. We use a deconvolved \tiger\
datacube to probe its kinematics at a resolution of about 0\farcs35
FWHM. Strong departures from circular motions are observed, as well as 
high velocity dispersion values on the inner side of the arms.  We interpret
the observed gas morphology and kinematics as the signature of streaming gas flows driven by a
$\sim 540$~pc diameter bar with $\Omega_p = 700 \pm 100$~\kms\ kpc$^{-1}$. This 
hypothesis is strongly supported by the predictions of a density wave model. 
This model predicts that the bar should lie at about 35\degr\ 
from the line of nodes, and implies gas inflow towards the central 
$\sim 50$~pc. The quadrupole pertubation due to
this bar is estimated to represent less than 2\% of the underlying
gravitational potential (a maximum torque of about 10\%), explaining the lack of
a direct detection via broad-band photometry in the visible. 
Despite its weakness, the inner bar of NGC~2974 may be
able to drive some gas within a 10~pc radius. We suggest
that the presence of such inner bars might be more common among early-type disk
galaxies than is generally thought, 
and that deep high-resolution emission-line imagery may be the best way to detect
such structures. 
\end{abstract}

\section{Introduction}

It is now evident that early-type galaxies are far from
being the simple, (violently) relaxed, isothermal, purely stellar systems
anticipated by the traditional picture developed by Hubble. They are often
found to contain a complex interstellar medium (ISM), stellar discs and/or
cusps and other discrete dynamical components (e.g. de Zeeuw et al. 2002). 
It now seems highly plausible
that much of this complexity originates in mergers or other interactions
between these galaxies and their environments. In addition, the statistical
frequency of active galactic nuclei (hereafter AGNs) in the form of
quasars and radio galaxies (which are generally associated with
elliptical galaxies) at medium to high redshift has led to the realization
that the majority of nearby elliptical galaxies may well harbor a central massive
black hole (e.g., Chokshi \& Turner 1992). 

Indeed, recent stellar kinematic studies using {\it Hubble Space Telescope
  (HST)\/} spectroscopy have provided evidence for the presence of black
holes in tens of early-type galaxies (Ferrarese \&
Merritt 2000; Gebhardt et al.\ 2000). They reveal the presence of a correlation
between black hole mass and bulge velocity dispersion, which suggests a causal
connection between the formation history of the black hole and that of the host
galaxy. 

One long-standing problem associated with the growth of black holes is the 
fueling mechanism. It has remained generally unclear how to transfer 
significant amounts of mass from the central kiloparsec of galaxy into its inner
regions (at a scale of 10~pc). While there are several known mechanisms that
can be invoked to 
initiate such gas inflow, most of them are commonly thought to be primarily
relevant to spiral galaxies rather than early-type galaxies. 

One such mechanism is density waves, the presence of which can be inferred from structures
like spirals or bars. These can develop naturally in galactic discs or be
initiated by `minor merger' of e.g. a massive spiral and a small satellite galaxy
(e.g., Pfenniger \& Norman 1990; Pfenniger 1991; Hernquist \& Mihos 1995). 
In particular, inner `secondary' bars in spiral galaxies (often referred to as
`bars within bars') have been identified as potentially important
agents to funnel gas into the innermost tens of parsecs (e.g., Shlosman, Frank
\& Begelman 1989; Erwin \& Sparke 1999; Emsellem et al.\ 2001). In this
scenario, a large-scale bar transports material into a kiloparsec-scale disc.
It may then be relayed by an inner bar that is able to transport the gas to
within  about 10 pc of the galactic nucleus, approximately where the 
central supermassive black hole's potential can take over. 
Structures with a radial extent of 240\,--\,750~pc, interpreted as inner bars,
are surprisingly common in early-type spirals (at least 25\% of barred S0/Sa
galaxies,  and maybe as much as 40\%, Erwin \& Sparke 2002). Such small 
bars have sometimes been linked to the central activity, as in e.g. NGC~6946 where
Elmegreen, Chromey \& Santos (1998) mention the possible role of the 210~pc
bar-like structure and the central starburst. 
There is evidence that large-scale bars are responsible for an increase  
in the central gas concentration (Sakamoto et al.\ 1999). It is however
important to emphasize 
that there is presently no strong direct evidence for the role of inner bars
in gas fueling within the central $\sim$\,50~pc. 
        
Martini \& Pogge (1999) analysed visible and near-infrared
{\it HST\/} images of 24 Seyfert 2 (spiral) galaxies and found that 20 of them
exhibit spiral arms of dust within the inner few hundred parsecs. These
inner spirals were thus suggested to play a crucial role in feeding gas into the
central engines (see also Regan \& Mulchaey 1999). The formation of
such small gaseous spiral structures may involve acoustic
instabilities 
in non self-gravitating discs (Elmegreen et al.\ 1998 and references therein). 
However, the arm-interarm contrast of such spiral arms is typically less than 0.1 mag, 
and it is not yet clear how efficiently these spirals can drive gas towards the
nucleus. 
A more important issue is that some of these spirals anyway require an external
driver to exist, such as a tumbling potential. Emsellem \& Ferruit (2000) reported on
the discovery of such an inner gaseous spiral in M\,104 (the Sombrero galaxy, a
highly bulge-dominated Sa spiral galaxy):\ {\it HST\/} imaging and
integral-field spectroscopy hinted towards the presence of  a (small) tumbling
bar potential.  

Since most radio-loud AGNs are found in early-type (E and S0) galaxies, we
started a 
{\it HST\/} program to study whether such inner spiral structures may also be
present in early-type galaxies. We are targeting galaxies from the sample of
Goudfrooij et al.\ (1994a, 1994b), who performed a ground-based imaging and spectroscopic 
survey of ionized gas in a complete, magnitude-limited sample of 56 elliptical
galaxies. We selected galaxies containing ionized gas features whose morphology and
kinematics showed promising in terms of hosting contrasted nuclear
gaseous spirals. In this paper we report on the results of {\it HST\/} imaging
and integral-field spectroscopy of the E3 galaxy NGC~2974, where we discovered
a strong inner trailing gas spiral.  

We first present the original datasets, which includes {\it
HST}/WFPC2 imaging and \tiger\ integral-field spectroscopy
(Sect.~\ref{sec:obs}). We then proceed in Sect.~\ref{sec:tigermaps} with the
two-dimensional maps resulting from a detailed analysis of the \tiger\ 
datacubes, and including a datacube deconvolution achieved with the use
of the narrow-band {\it HST\/} imaging as a guide. The 
dynamical modeling conducted to fit the observed stellar and gaseous
kinematics is presented in Sect.~\ref{sec:model}. A brief discussion is provided
in Sect.~\ref{sec:disc}, followed by concluding remarks in Sect.~\ref{sec:conc}. 

In this paper, we will assume a distance of 21.5~Mpc (from surface brightness
fluctuations, worked out by Tonry et al.\ 2001), and use the variable $d_{21.5} =
D[\mbox{Mpc}] / 21.5$ to normalize all quantities that are distance dependent.
We favour this to a normalization like (e.g.) $h_{75}$ (see Emsellem
\etal 1996 for a rationale of this choice). This yields a scale of $\sim 104$~pc/arcsec.

\section{Observations and data reduction}
\label{sec:obs}

\subsection{\tiger\ two-dimensional spectroscopy}
\label{sec:tigerobs}

\subsubsection{Data reduction}
\label{sec:tigdata}

We obtained integral field spectroscopy of NGC~2974 during
a run in November 1993 at the Canada-France-Hawaii Telescope using the
\tiger\ instrument.
This spectrograph makes use of an array of about 500 micro-lenses
whose diameter was set to $0\farcs39$ for these observations.
Two configurations were used to cover 
the spectral domains around the Mg\lda5172 triplet (\blue\ hereafter) and 
the \Ha, \NII\lda6548,\lda6583 and \SII\lda6717,\lda6731
emission lines (\red\ hereafter). The characteristics of both \red\ and \blue\ 
configurations are described in Table~\ref{tab:tiger}.
\begin{table}
\caption[]{Observational characteristics of the \tiger\ spectrographic 
exposures. The fields of view are only indicative as the merged \tiger\ 
exposures do not cover a rectangular area.}
\begin{center}
\begin{tabular}{@{}lll@{}}
\hline \hline
\multicolumn{3}{c}{~~} \\ [-1.8ex]
Nov 1993 Run & \Blue\  & \Red\  \\ [0.5ex] 
\hline
Lens diameter & $0\farcs39$ & $0\farcs39$ \\
Final spatial sampling & $0\farcs36$ & $0\farcs36$ \\
Field of view & $8\farcs4 \times 7\farcs8$ &  $7\farcs4 \times 7\farcs8$ \\
\# of exposures & 3 & 2 \\
Total exp. time & 135 mn & 105 mn \\
\# of merged spectra & 382 & 429 \\
Spectral sampling & 1.5 \AA\,$\mbox{pixel}^{-1}$ & 
 1.5 \Ag$\mbox{pixel}^{-1}$ \\ 
Spectral resolution & 4.0 \AA\ (FWHM) & 3.2 \AA\ (FWHM) \\
Spectral domain & 5090 - 5580 \Ag & 6550 - 6960 \Ag \\
Seeing (FWHM) & $1\farcs2$ & $0\farcs85$ \\
Stellar template & HR 1681 \rlap{(K0III)} & \\
\hline
\end{tabular}
\end{center}
\label{tab:tiger}
\end{table}

The reduction of these data was achieved with the \tiger\ software package
written at the Lyon Observatory (Rousset 1992). This included bias subtraction, 
pixel-to-pixel flat fielding, spectral extraction, wavelength
calibration, flat fielding, removal of cosmic rays,
differential atmospheric refraction correction and flux calibration for all individual
exposures. The resulting data-cubes were then resampled to a square grid
and merged. 
Details about these reduction steps can be found in Emsellem \etal (1996).
In the next paragraph, we only mention points specific to the
Nov.\ 1993 \tiger\ run and the NGC~2974 data.

A slightly incorrect position angle of the CCD with respect to the
lens array led to some unwanted contaminations between adjacent spectra
at their edges: this only affected the \red\ configuration, for which
spectra were safely truncated to a wavelength range of 6570--6960 \AA.
The wavelength calibration was found to be accurate to within
0.03 pixel: this was made possible through the use of a Fabry-Perot etalon
which provides regularly spaced Airy lines.
The absolute flux calibration of the \red\ spectra was performed 
using spectro-photometric standard stars. For the \blue\ configuration we used 
an available flux-calibrated long-slit spectrum (Goudfrooij \& Emsellem
1996).

The stellar kinematics were derived from the \blue\ spectra
via a slightly modified version of Bender's (1990) Fourier Correlation
Quotient (FCQ hereafter) method. Line strengths were measured
using an empirical correction for the dispersion as in Emsellem et al.\
(1996). Since no specific attempt was made to calibrate our data onto the
Lick system, we remind the reader that systematic offsets may exist between
our line-strength values and other published ones. 
The emission-lines in the spectra were fitted by Gaussian profiles using the
FIT/SPECTRA software written by Arlette Rousset (Lyon Observatory - Rousset 1992).  

\subsubsection{Stellar template subtraction}
\label{sec:contsub}

\paragraph*{(a) The \blue\ spectra: the \NI\lda5200 line} ~

\par\smallskip\noindent
As discussed in Goudfrooij \& Emsellem (1996), when ionized gas is present
the \NI\ emission-line doublet at \lda\lda 5198,5200 (hereafter \NI\lda5200) can
jeopardize the analysis of stellar dynamics and stellar populations of galaxies in the
spectral domain around the \mgb\ feature . 
We therefore devised an iterative procedure to eliminate the contribution 
of \NI\lda5200 to our \blue\ spectra. This method will be described and
illustrated in detail in a separate paper, so we only provide an outline
here:  
\begin{enumerate}
\item We first derive the stellar kinematics with FCQ and the original
\blue\ spectra as they are.
\item We then use a library of pure absorption-line spectra, including
template stars as well as galaxies devoid of emission lines, to fit each of
our \blue\ spectra, after proper broadening and redshifting (using the result
of step (i)).
A small spectral region around the expected location of the \NI\lda5200 line
was masked out during this fitting process. 
\item We then subtract the resulting model absorption-line spectra
and make a fit of the \NI\lda5200 system on the
residual pure emission-line spectra.  
An example is presented in Fig.~\ref{fig:NIfit}.
\item The fitted \NI\lda5200 line systems are then subtracted
from the original \blue\ spectra, thus resulting in spectra
from which the \NI\lda5200 contribution is eliminated.
\item Finally, we re-measure the velocities $v$ and dispersions $\sigma$
from the ``cleaned'' spectra: these values are the ones adopted
for the remainder of the paper.
\end{enumerate}

\figstart{figure=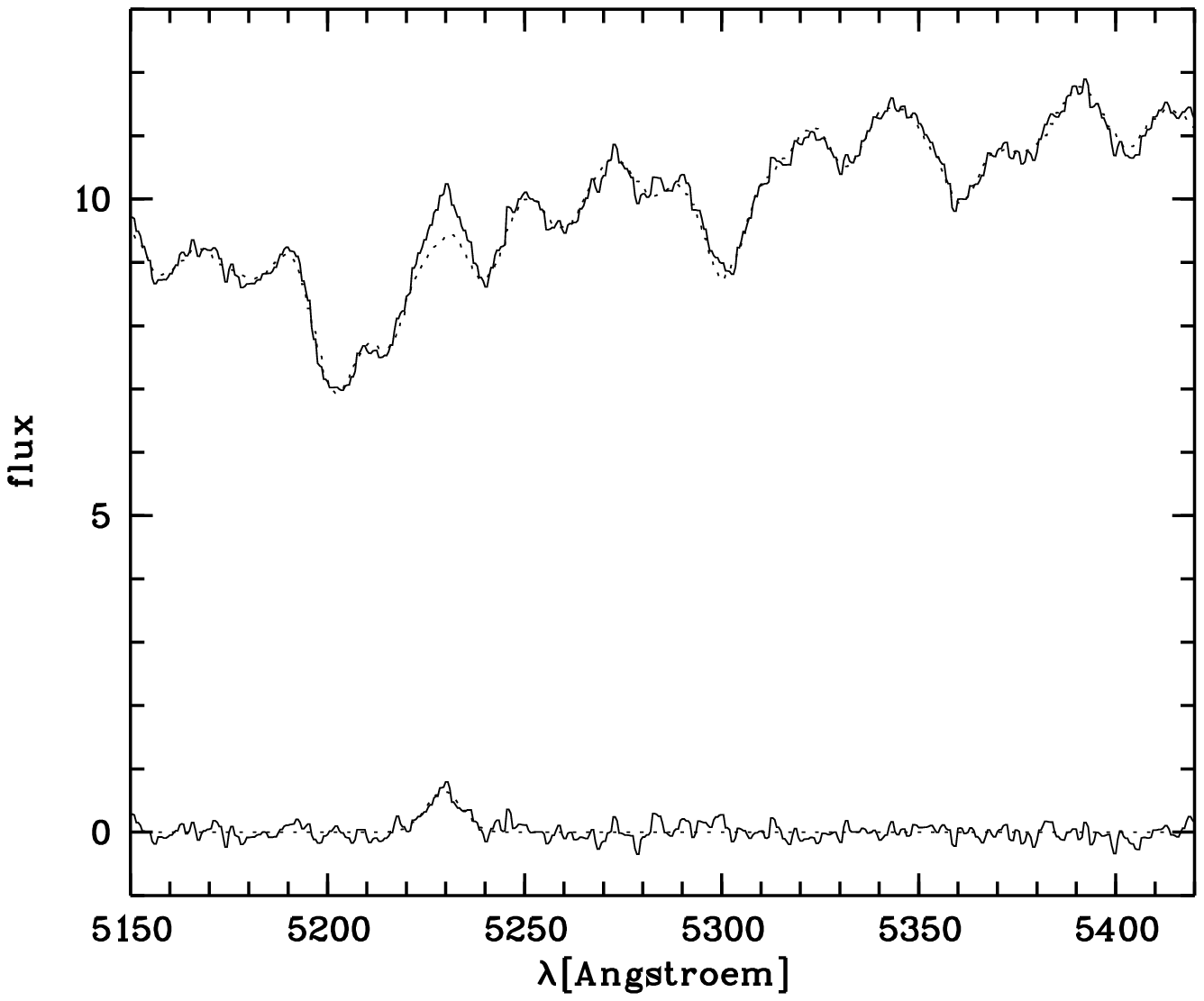,width=\hssize}
\caption[]{Example of the stellar template subtraction method 
for the \blue\ spectra. The spectrum at the top
shows the original spectrum (solid line) and best fit using a library of
absorption-line template spectra (dotted line). At the bottom, the residual
spectrum (solid line) and the corresponding \NI\lda5200 line fit (dotted
line) is presented. Flux unit is \ergcmsaa.}
\label{fig:NIfit}
\figend

\paragraph*{(b) The \red\ spectra} ~

\par\noindent
Just as described above, we use a library of star and galaxy spectra to remove
the stellar continuum in the \red\ spectra. However, the limited wavelength
coverage and the high equivalent width of the emission lines in this spectral
domain makes this a somewhat difficult
task. For that reason, we built an optimal template using the stellar kinematics
derived from the \blue\ spectra and only including the stellar templates
which were used in the fit of the \blue\ spectra: this prevents artificial
variations of (e.g.) the width of the \Ha\ absorption line. An example of
such a fit and subtraction is shown in Fig.~\ref{fig:fitNII}. The residual 
`pure' emission-line spectra were then fitted using the {\sc fit/spectra} software
package (Rousset 1992) assuming Gaussian profiles for each individual line. 

\figstart{figure=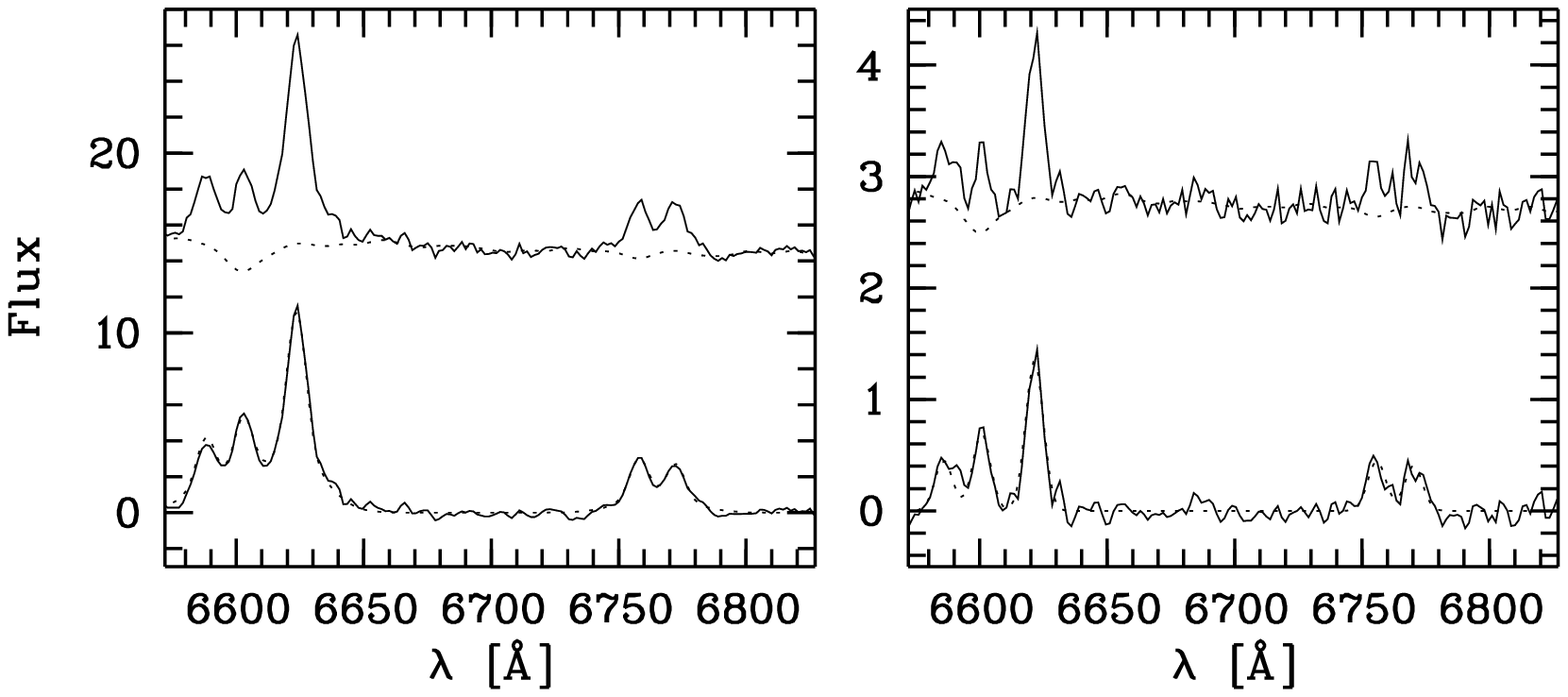,width=\hssize}
\caption[]{Similar plots as in Fig.~\ref{fig:NIfit}, 
now for the \red\ spectra. Two examples
are shown, each time with the original spectrum (solid lines, 
top) and their respective fit (dotted lines, top),
as well as the residual spectra (solid lines, bottom) and 
the Gaussian fit to the emission lines (dotted line, bottom).
Flux unit is \ergcmsaa.}
\label{fig:fitNII}
\figend

\subsection{HST/WFPC2 imaging}
\label{sec:HSTimaging}
 
We used the Wide Field and Planetary Camera 2 (WFPC2) aboard {\it HST\/} to
obtain high spatial resolution images of stars and ionized gas in the inner
regions of NGC~2974 on April 16-17, 1997. 
Three exposures were taken with the Linear Ramp Filter 
FR680 at a filter wheel rotation of +15\degr. This filter configuration
results in a passband covering the \Ha\,\lda6563 + \NII\,\lda\lda6548,6583
emission lines at the redshift of NGC~2974 when observed at the chosen
position on the PC chip of WFPC2. Two exposures were also obtained with each
of the filters F547M and F814W, whose passbands are free from (strong)
emission lines at the redshift of NGC~2974. Total exposure times were 1400
s, 500 s, and 5100 s for the F547M, F814W, and FR680 filters,
respectively. For each filter, one of the images was spatially offset by
0\farcs5 from the others, which corresponds to an approximately integer
pixel shift in both PC and WF CCDs. This was done to enable the
identification of hot pixels during image combination. After standard
pipeline processing and alignment of the images, we combined the images per
filter using the STSDAS task {\sc crrej}, which effectively removed both
cosmic rays and hot pixels. Flatfielding of the FR680 images was done 
using a reference flatfield taken through the F658N filter.

Since NGC~2974 contains prominent dust structures near its nucleus (see
below and Fig.~\ref{fig:HSTimages}), the removal of the continuum
contribution from the FR680 image is more complicated than simply
subtracting a F547M or F814W image. First of all, the F547M and F814W images
were aligned with the FR680 image. The F547M and F814W images were then
matched to the same PSF.  As no bright stars are present on the WFPC2
frames, this was done by building PSFs for all three filters using {\sc
tinytim} (Krist 1992), and convolving each filtered image with the PSF of
the other filtered image. A similar procedure was performed to match the PSF
of the FR680 image to that of the F547M and F814W images. 
A synthetic image of the continuum at 6620 \AA\ was then constructed with
the {\sf calcphot} task in the {\sc synphot} package of STSDAS. We used the
F547M and the ratio of the F547M and F814W images in conjunction with the
assumptions that {\it (i)\/} the unreddened spectral 
energy distribution of the stellar population matches that of the 
``E2'' early-type galaxy spectrum of Bica (1988), and {\it (ii)\/} the
dust features in NGC~2974 cause reddening that is described by the
Galactic interstellar extinction curve of Rieke \& Lebofsky (1985), taking
into account that half of the stellar light from NGC~2974 is in front of the
nuclear dust features. This process will be described in more detail in
Paper II of this series (Goudfrooij \& Emsellem 2003, in preparation). This 
synthetic 6620 \AA\ continuum image was subtracted from the FR680 image,
yielding the \Ha+\NII\ emission-line image. A $V\!-\!I$ colour map was also
constructed, using the F547M and F814W images. Figure~\ref{fig:HSTimages}
shows grey-scale representations of the F547M, F814W, $V\!-\!I$, and
\Ha+\NII\ images. 

These images will be discussed in detail in 
Goudfrooij \& Emsellem (2003, in preparation) but will be used here
for a datacube deconvolution (Sect.~\ref{sec:deconv}) and in the modelling
of the gaseous component (Sect.~\ref{sec:wave}).

\lfigstart{figure=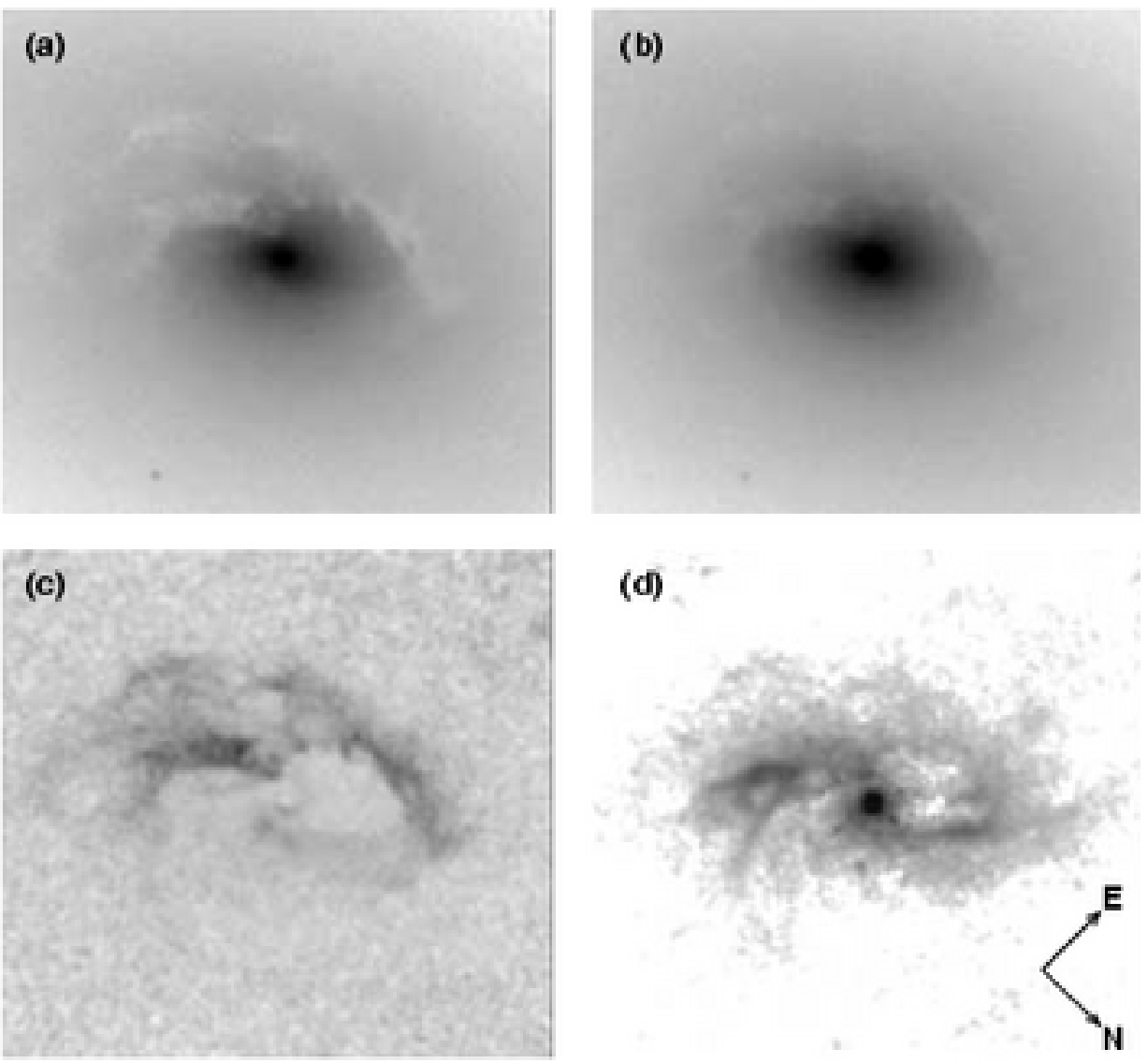,width=\hdsize}
\caption[]{{\it HST}/WFPC2 images of the central 6$\times$6 arcsec$^2$ of 
  NGC~2974. with a resolution of $\sim$\,0\farcs1 ($\cor 10.4\;$\dist~
  pc),  except for the \Ha+\NII\ image in {\sf (d)} which has resolution
  $\sim$\,0\farcs13.  The orientation on the sky is indicated in panel {\sf
  (d)}. {\sf (a)} The F547M (5454~\AA) image, and {\sf (b)} the F814W
  (8269~\AA) image. The greyscales in both {\sf (a)} and {\sf (b)} use 
  a logarithmic stretch. 
  Darker shades represent brighter regions. {\sf (c)} The
  $V\!-\!I$ colour map, obtained from the ratio of {\sf (a)}  to {\sf (b)}
  as explained in Sect.~\ref{sec:HSTimaging}. The greyscales range
  linearly from of 0.84 to 1.64 mag in $V\!-\!I$. Darker shades
  represent redder colours. {\sf (d)} The H$\alpha$+\NII\ image. The
  greyscales (in units of erg cm$^{-2}$ s$^{-1}$ arcsec$^{-2}$) range 
  logarithmically from $2.0 \times 10^{-15}$ to $4.0 \times 10^{-13}$,
  while the peak intensity is $9.6 \times 10^{-13}$. Darker shades represent
  brighter regions.
}
\label{fig:HSTimages}
\lfigend

\section{Results from the \tiger\ datacubes}
\label{sec:tigermaps}

In the following Sections, we present the maps of the morphology
and kinematics of the stellar and ionized gas components, as
well as the \mgb, Fe5270 and Fe5335 line-strength maps.

After barycentric correction, the systemic velocity of NGC~2974 was measured
from the stellar absorption lines and found to be $1888 \pm 10$~\kms,
significantly lower than the commonly used value of 1924~\kms\ (Davies et al.\
1987), but perfectly consistent with the value of $1890 \pm 40$~\kms\ quoted
by Kim \etal (1988). All maps are oriented with the major axis along the
horizontal axis. We remeasured the position angle of the major axis in the
outer part of the galaxy on direct \tiger\ images and found PA$=42.2\degr$
(the North axis will then be always 132.2 degrees from the vertical axis). 
Note that there is a slight isophote twist of about 4~degrees
towards the centre, partly due to dust extinction which is most
prominent in the southeast side of the galaxy (cf.\
Fig.~\ref{fig:HSTimages}), thus defining its near side.   

In order to compare our results with previously published kinematics of
NGC~2974 (using long-slit spectra), we simulated long-slit
cuts through our  data cubes following characteristics of the observations
made by Bender et al.\ (1994) and Cinzano \& van der Marel (1994; hereafter
CvdM94) for the stellar component, and by Zeilinger et al.\ (1996) for the
ionized gas component. Comparisons are plotted in Fig.~\ref{fig:starcomp}
and Fig.~\ref{fig:gascomp} for the stellar and ionized gas components
respectively, after subtracting the systemic velocity\footnote{We applied a  
barycentric correction to the Zeilinger et al.\ data sets taking
into account the dates of their observations.}.
Both comparisons show excellent agreement. 

\figstart{figure=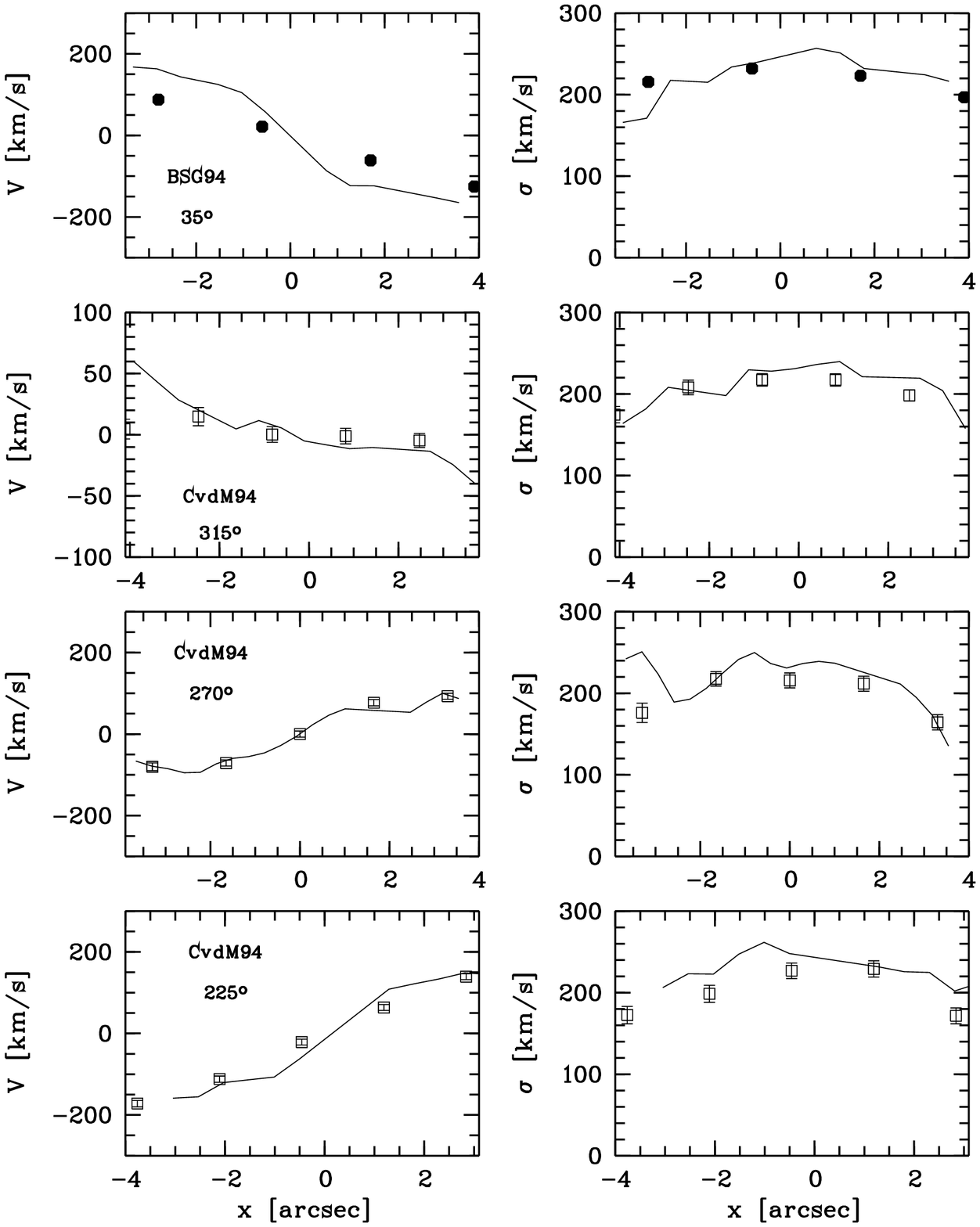,width=\hssize}
\caption[]{Comparison between the \tiger\ stellar kinematics 
(solid lines) for NGC~2974 with published data. Velocity $V$ is shown in the
left panels, and dispersion $\sigma$ in the right panels. Open squares are
from CvdM94, and filled squares are from BSG94. 
Velocities are relative to the systemic velocity of NGC~2974.}
\label{fig:starcomp}
\figend
\figstart{figure=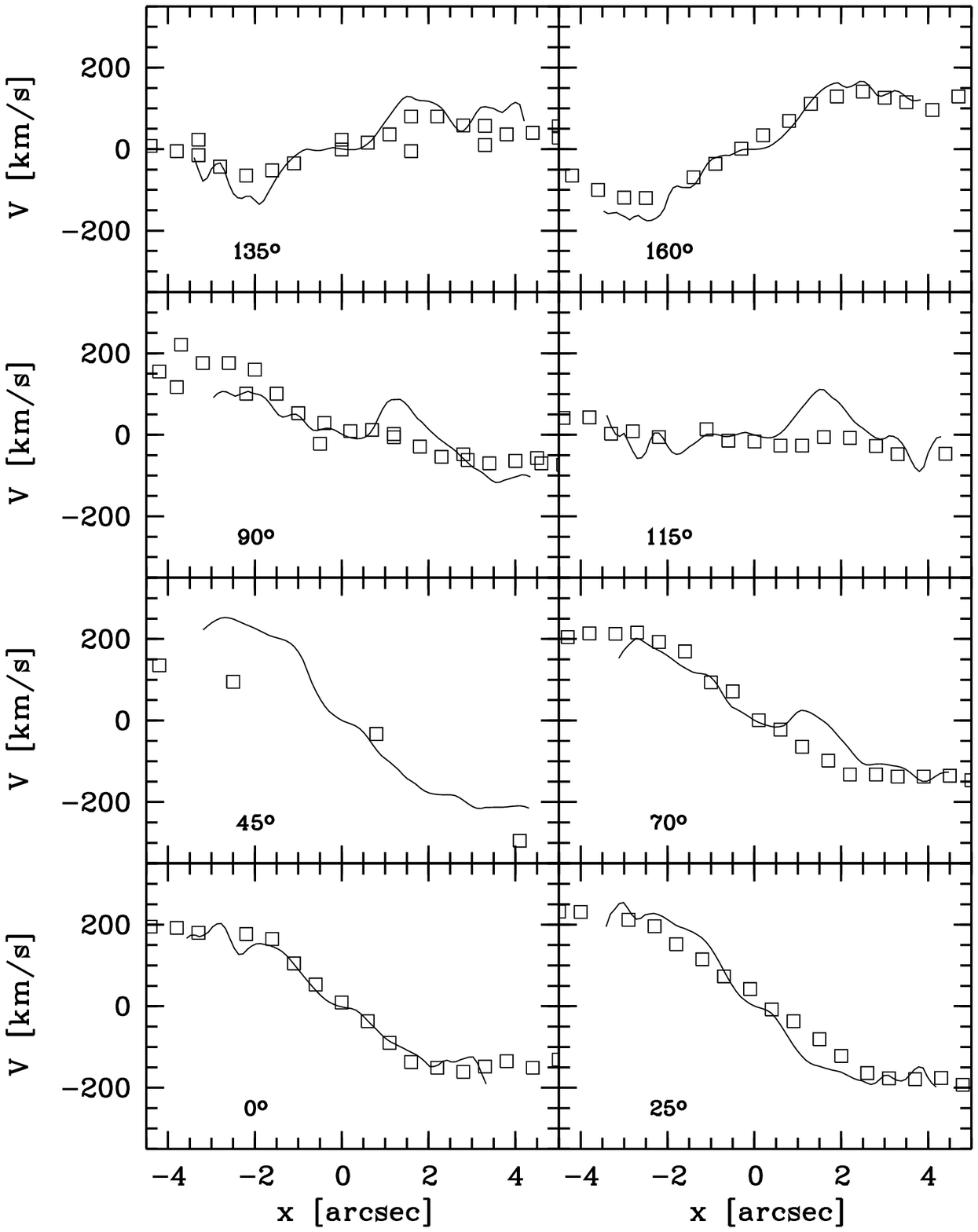,width=\hssize}
\caption[]{Comparison between the \tiger\ gas kinematics 
(solid line; $V_g$ only) and the published long-slit
kinematics (open squares) presented by Zeilinger et al.\ 
(1996). }
\label{fig:gascomp}
\figend

\subsection{The stellar component}

The maps in this Section have all been reconstructed using the sky sampling of
$0\farcs36$ for the pixel size: this way, one pixel in a map corresponds to one
spatial element in the \blue\ datacube. 

\subsubsection{Stellar kinematics}

In Fig.~\ref{fig:IVSmap}, we present the stellar velocity $V$ and dispersion
$\sigma$ \tiger\ maps. 
The observed velocity field exhibits some departures from 
what is expected from axisymmetry. We measure a tilt of the zero-velocity 
curve of $9\deg \pm3\deg$. This might partly be due to dust
extinction perturbing the line-of-sight kinematics, although extinction
is significant only on the South-East side. This
should be confirmed by spectroscopy with higher signal-to-noise ratio. 
The central velocity gradient is about 110~\kms\ arcsec$^{-1}$
(or $\sim 1094$~\dist$^{-1}$\ \kms\ kpc$^{-1}$).
The velocity dispersion field is nearly flat in the central two
arcseconds with a value around $\sim 215$~\kms, and slowly decreases
outwards. 
\figstart{figure=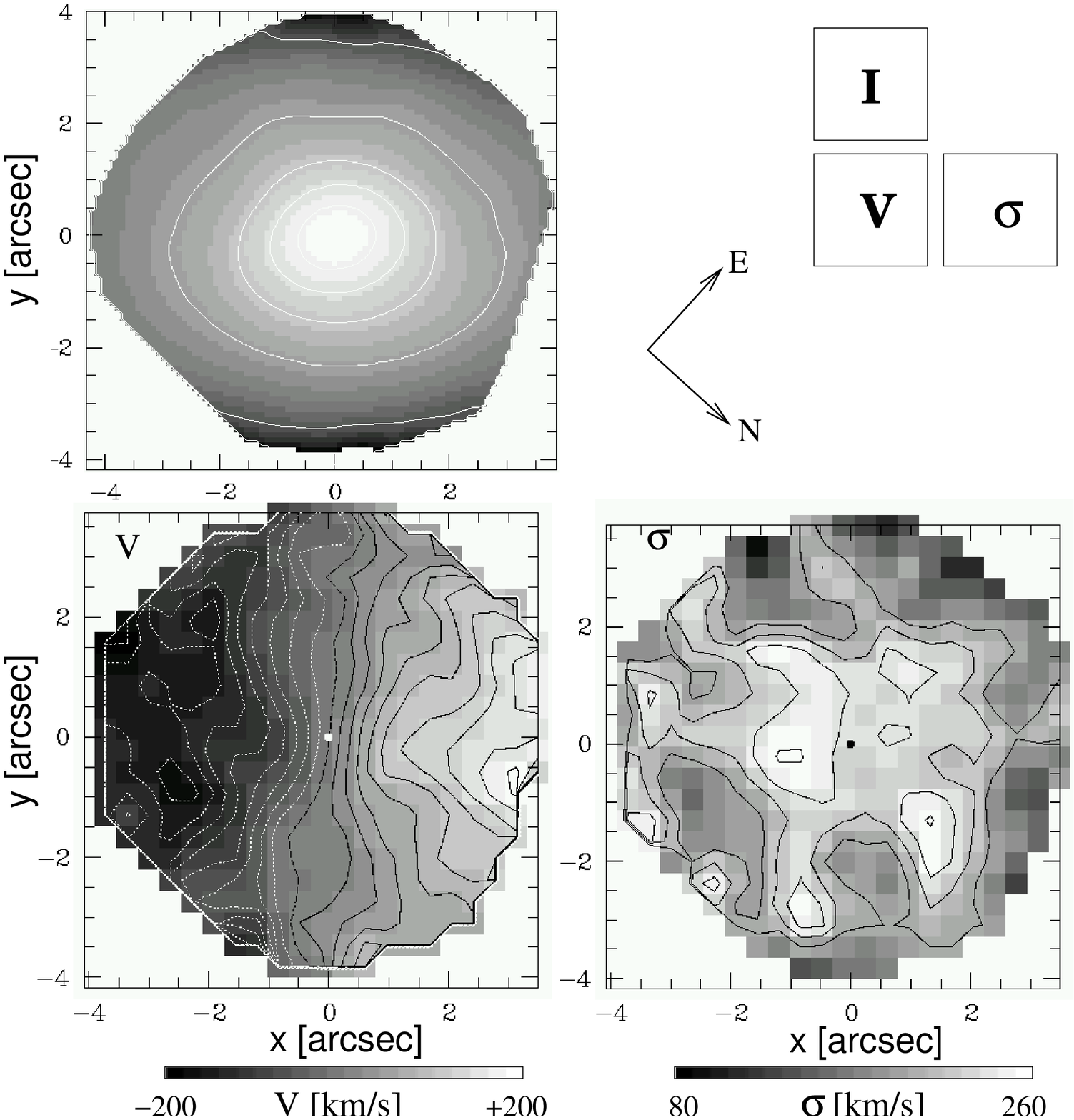,width=\hssize}
\caption[]{\tiger\ stellar kinematics of NGC~2974: surface brightness
(top left), velocity (bottom left) and velocity dispersion (bottom right)
maps.  
The step of the surface brightness contours is 0.5 mag~arcsec$^{-2}$,
the velocity contour step being 20~\kms\ for both the velocity and
dispersion. The central dots in each figure mark the centre of the
isophotes. North has been orientated at 132.2 degrees from the vertical axis
as indicated.} 
\label{fig:IVSmap}
\figend

\subsubsection{Line strengths}

We derived line strength maps from the \blue\ \tiger\ spectra, 
namely \mgb, Fe5270, Fe5335 and Fe5406. Here we present the \Fem\ map,
(classically defined as $(\mbox{Fe5270} + \mbox{Fe5335}) / 2$) as well as 
the \mgb\ map {\em before\/} and {\em after\/} correction
for the contribution of the \NI\lda5200 emission line
(Figs.~\ref{fig:Mgmap} and \ref{fig:Femap}), to illustrate its
effect on the \mgb\ line strength (see Goudfrooij \& Emsellem 1996). 

The highest \mgb\ value of $5.1 \pm 0.2$ is reached about $-0.9$~arcsec 
from the centre along the major-axis, with its symmetric point having
\mgb\ of $4.4 \pm 0.2$ (the central value is $4.3 \pm 0.2$). There is a slight decrease of \mgb\ 
outwards, with values at the edge of the field being around 4.
The \Fem\ and Fe5406 maps are consistent with being flat. 
Higher signal-to-noise spectra would help confirming the absence
of significant line strength gradients in the central arcseconds.

\figstart{figure=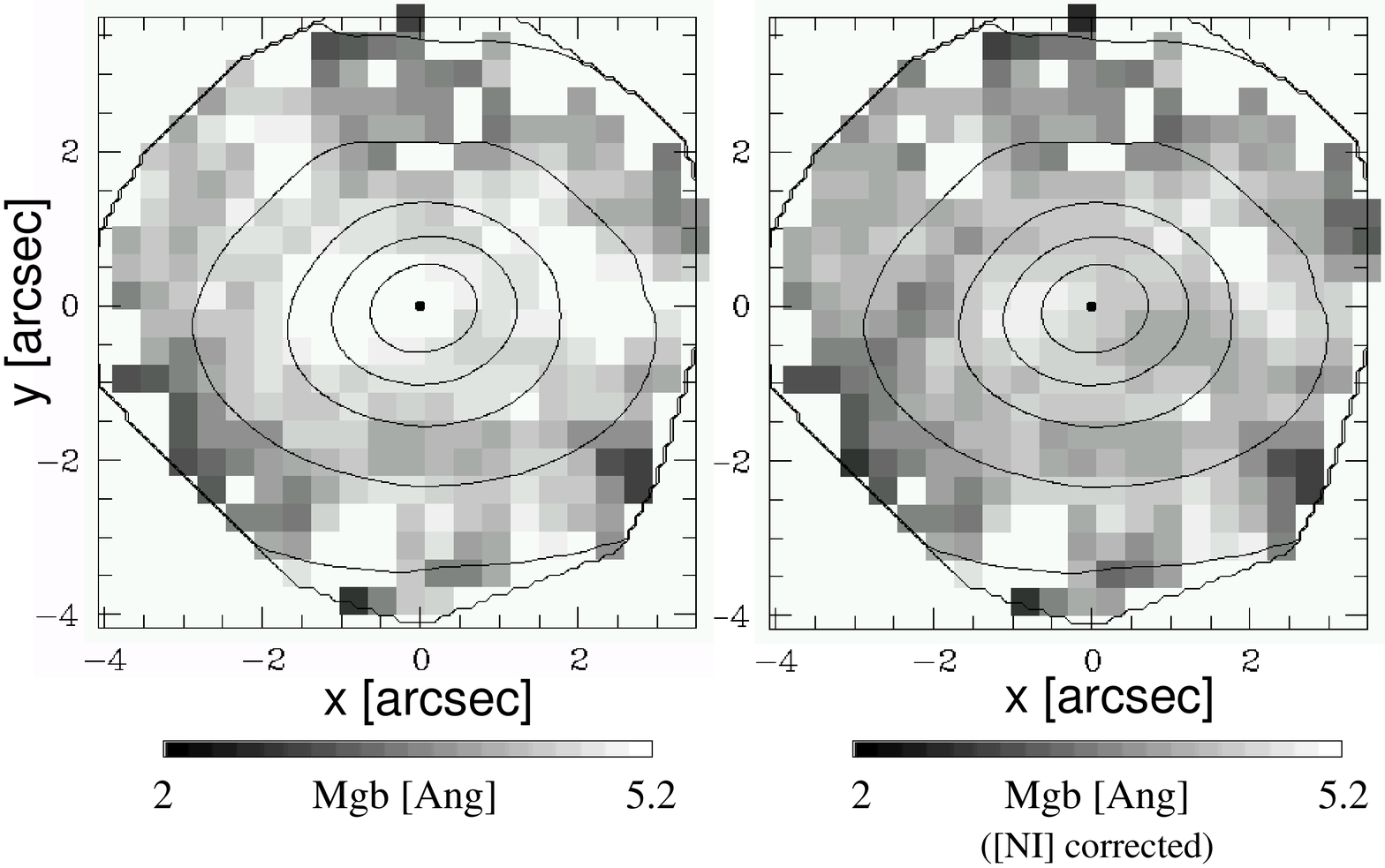,width=\hssize}
\caption[]{\mgb\ line strength maps before (left panel) 
and after (right panel) correction for the [NI]\lda5200 
emission line. Surface brightness contours from the \blue\ \tiger\ data 
(cf.\ Fig.~\ref{fig:IVSmap}) are overplotted}
\label{fig:Mgmap}
\figend
\figstart{figure=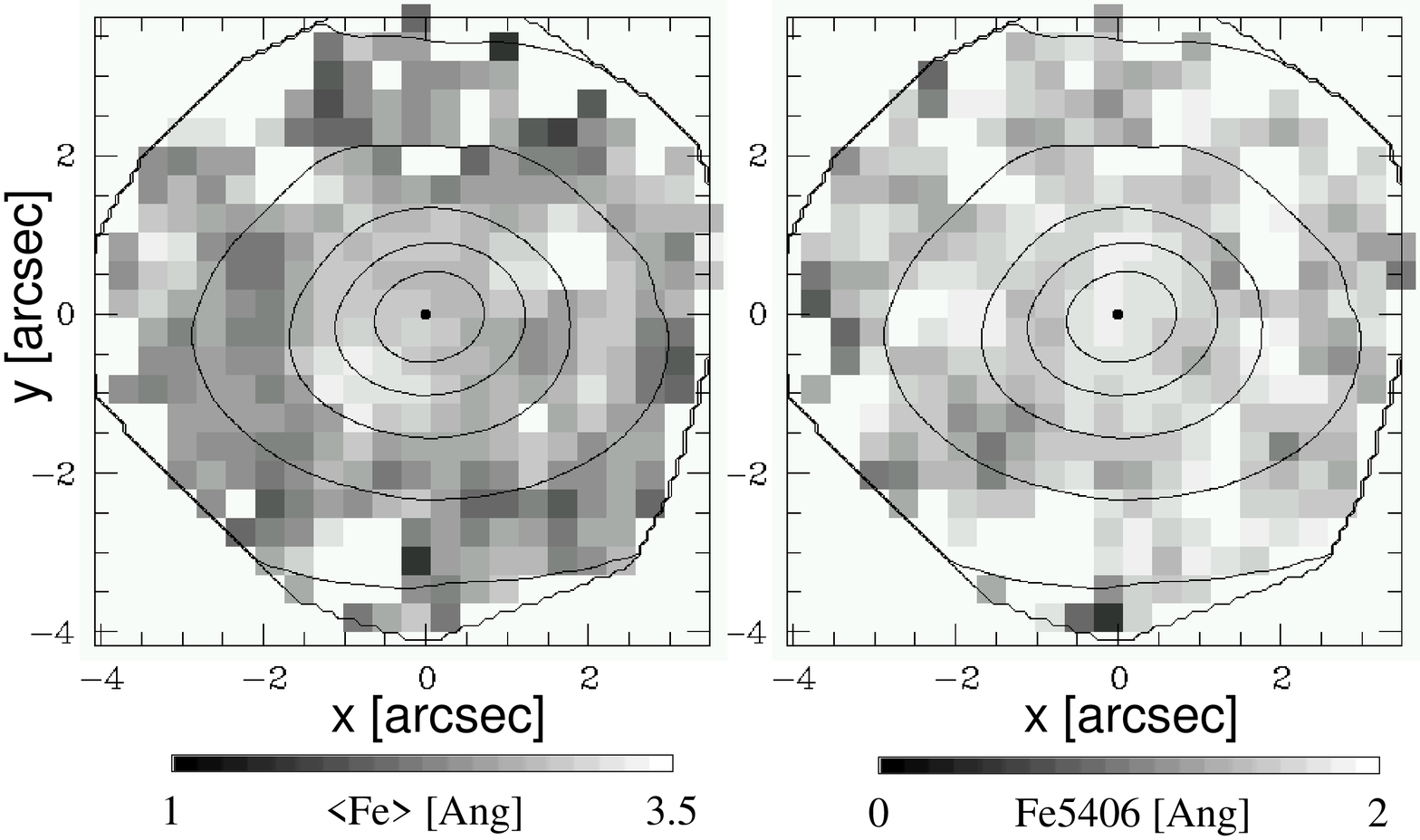,width=\hssize}
\caption[]{\Fem (left panel) and Fe5406 (right panel) 
line strength \tiger\ maps for NGC~2974. Contours as in
Fig.~\ref{fig:Mgmap}} 
\label{fig:Femap}
\figend

\subsection{The gaseous component}

We now present the emission-line maps derived from fits of the emission
lines present in the \red\ data cube. These maps were built 
using a pixel size of $0\farcs18$, half of the final sky sampling.

\subsubsection{Flux maps and spectra}
\label{sec:gaspec}

We detect \Ha\ as well as the \NII\ and \SII\ doublets throughout the whole
field of view. \Ha\ is in fact observed on a much larger scale, as nicely emphasized
with the maps produced by Plana \etal (1998). From our \tiger\ data,
we cut out regions where the intensity of the
\NII\lda6583 emission line was less than 5 times the 1~$\sigma$ noise level
to avoid spurious features in the maps. This of course implies a lower
flux threshold for the \SII\ lines which are significantly weaker than the
\NII\ lines. However, we kept the same field size for all lines to avoid any
confusion. 

Spectral profiles of the emission lines show large line widths in the
centre. This is true for both \Ha\ and the forbidden lines (\NII, \SII), so
that it is not due to the presence of a
Broad-Line Region of an active galactic nucleus (AGN). 
Spatial mapping of the ionized gas showing such broad lines shows that
it is spatially unresolved in the \tiger\ data. We therefore modelled it with
a fixed radial velocity  $v_{{\rm wide}}$ and dispersion $\sigma_{{\rm
 wide}}$. The values for $v_{\rm wide}$ and $\sigma_{\rm wide}$ 
were determined on the central spectrum, in which the wings have the best
contrast and signal-to-noise ratio: we find $v_{wide} = 1965$~\kms and
$\sigma_{wide}= 460$~\kms. Errors bars on these values are difficult to derive
since it should include both the formal error bar as well as the uncertainty
due to the noise in the data. We estimated them using the 25 spectra which are 
within 1\arcsec\ of the centre and were found to exhibit such a wide
component. These central 25 spectra were fitted with an additional Gaussian
component. and we then find rms uncertainties of 35~\kms and 61~\kms on
the velocity and velocity dispersion respectively. All other spectra were
fitted with a single Gaussian. 

The \NII\lda6583 emission-line flux map is presented in
Fig.~\ref{fig:NIIwithspec} with spectra illustrating the variations of line
profiles and line ratios over the field. Other maps (\SII\ and \Ha) are
presented in Fig.~\ref{fig:otherlines}. They all exhibit a two-arm spiral
structure that 
is consistent from one emission line to the other, with an additional bright
central peak. The SW arm (negative abscissa) is brighter than the NE one. In
the inner part of each arm, the emission-line profiles are complex. 
This is probably caused by the superposition of components at different
velocities. 

Note that a simple {\em a posteriori\/} check of the validity of the 
stellar continuum subtration method described in Sect.~\ref{sec:contsub} 
can be done by reconstructing the two-dimensional \NI\lda5200 intensity map  and
comparing it with the \NII\lda6583 map (Fig.~\ref{fig:otherlines})
(for which the surrounding continuum is essentially featureless).
The match is excellent (see Fig.~\ref{fig:NImap}), particularly when 
considering the low signal-to-noise ratio of the residual
\NI\lda5200 line spectra. 

The \SIIwa\ / \SIIwb\ ratio ranges from $\simeq$1 at the centre to
$\simeq$1.4-1.5 in the arms, corresponding to electron densities $N_e$ of
500~\cmc\ and less than 100~\cmc\ (low density limit), respectively.
The \NIIwb\ / \Ha\ ratio is roughly constant over the field of view,
with values between 2 and 2.2. This seems to contradict the measurements made by
Zeilinger et al. (1996), who reported \NIIwb\ / \Ha\ values as high as 8
at the centre, a factor of two higher than ours at the centre.
We suggest that this discrepancy is mainly due to the fact that Zeilinger et al. (1996)
did not flux-calibrate their spectra (which were taken for the purpose of
measuring gas kinematics) and did not correct them for the
underlying (H$\alpha$) stellar absorption. 

The flux distribution of the \SII\ lines is less centrally peaked
than that of the \NII\ lines, with the \SIIww\ / \NIIwb\ ratio ranging
from $\simeq$0.4 at the location of the nucleus to $\simeq$0.65 in the arms.
This is usually interpreted as a decrease of the \SII\ emission in the
dense nuclear regions, due to the lower critical electronic densities of the
\SII\ lines with respect to the \NII\ line doublet ($N_e^{crit} = 
1.9$ $\times$ \ten{3}, 1.2 $\times$ \ten{4} and 8 $\times$ \ten{4}~\cmc\ for 
the \SIIwa, \SIIwb\ and \NIIww\ lines, respectively).

The \Ha\ flux integrated over a 3\arcsec\ radius from the nucleus is 
$\simeq$ 1.9 $\times$ \ten{32}\ \dist$^2$~W, with a $\simeq$ 4.7
$\times$ \ten{31}\ \dist$^2$~W contribution from the central unresolved
peak. Assuming a uniform electron density $N_e$ = 100~\cmc\ and a pure hydrogen
gas, this corresponds to a mass of ionized gas in the arms of 6.8 $\times$
\ten{4}\ solar masses. For the central unresolved peak and assuming this time
an electron density of 500~\cmc, we obtain a mass of ionized gas of 4.4 $\times$
\ten{3}\ solar masses. Note that the quoted ionized gas masses are inversely
proportional to the value of $N_e$.

\lfigstart{figure=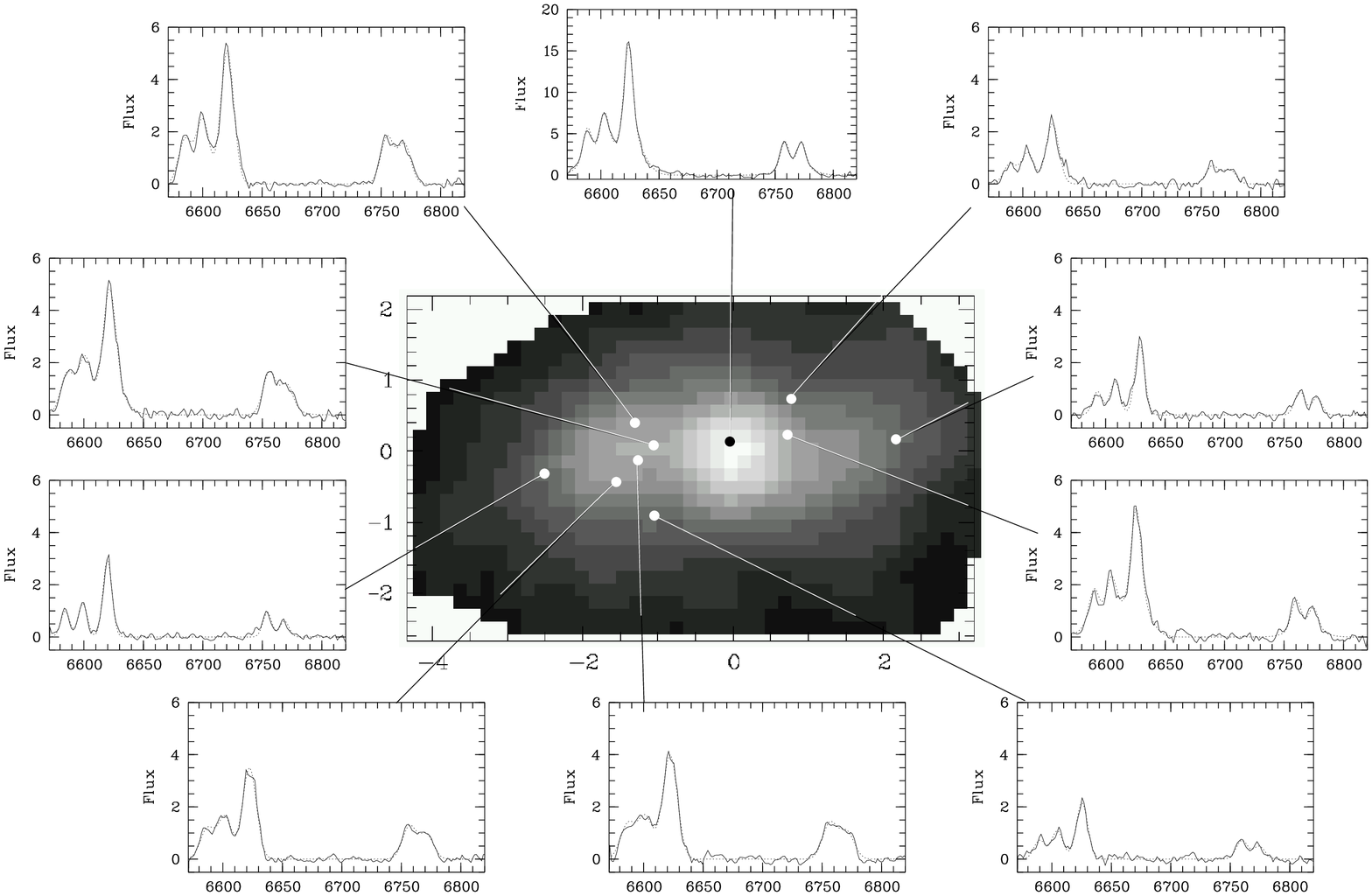,width=17.6cm}
\caption[]{Flux map of the [NII]\lda6583 emission line derived from the 
\red\ \tiger\ datacube.
Spectra (solid lines) and their corresponding multi-Gaussian fits (dashed
lines) are shown in various places over the field. Note that all flux scales
are the same, except for the central spectrum shown in the central top
panel. Flux unit is $10^{-16}$~\ergcmsa\ for the spectra.}  
\label{fig:NIIwithspec}
\lfigend
\lfigstart{figure=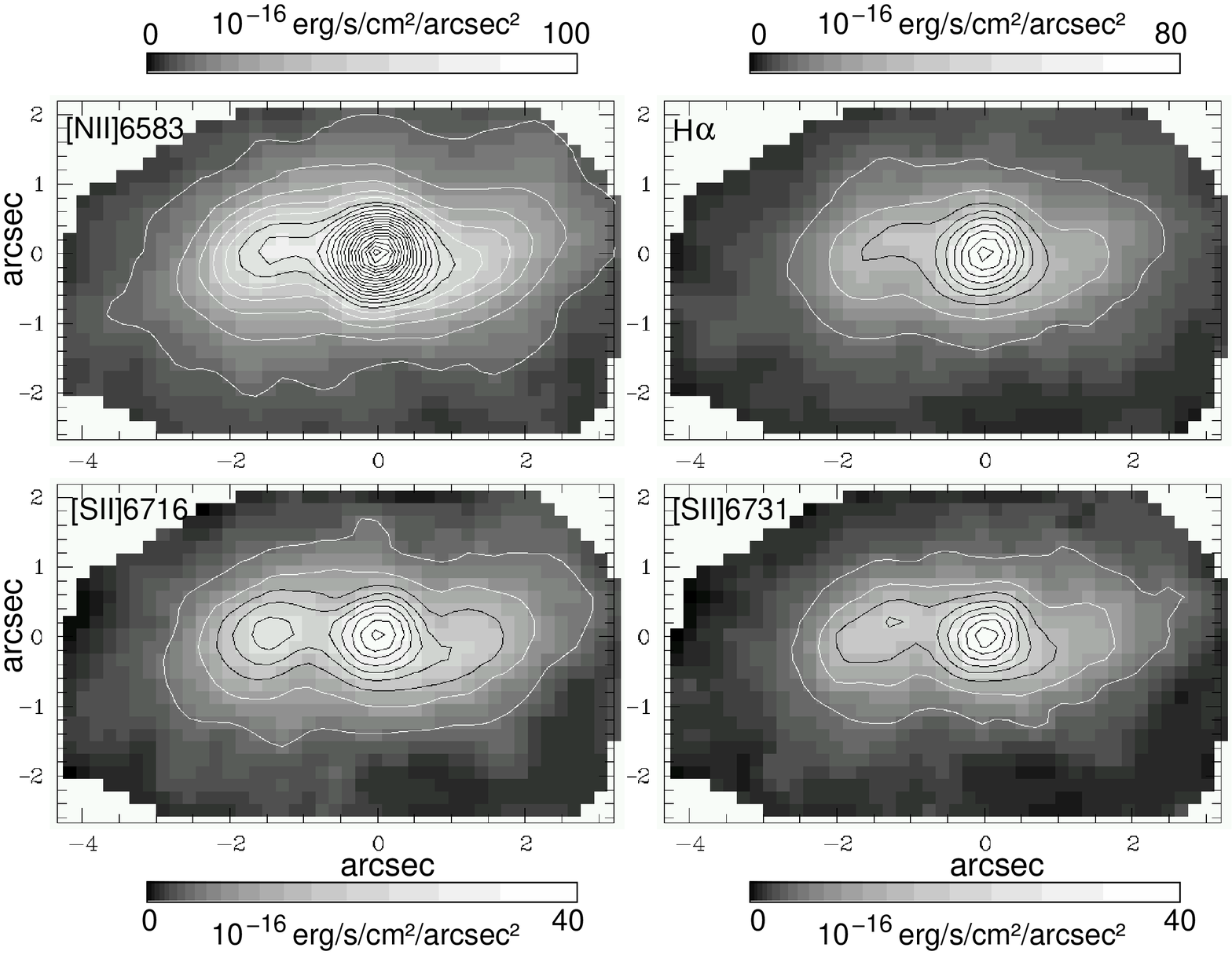,width=17.6cm}
\caption[]{Flux maps of \NII\lda6583 (top left), \Ha\ (top right),
\SII\lda6717 (bottom left) and \SII\lda6731 (bottom right). The lowest
isocontours corresponds to 10$^{-15}$~\ergcmsa, and the isocontour step
is 10$^{-15}$~\ergcmsa\ for \NII\ and \Ha, and $5\;10^{-16}$~\ergcmsa\ 
for both \SII\ maps.} 
\label{fig:otherlines}
\lfigend
\figstart{figure=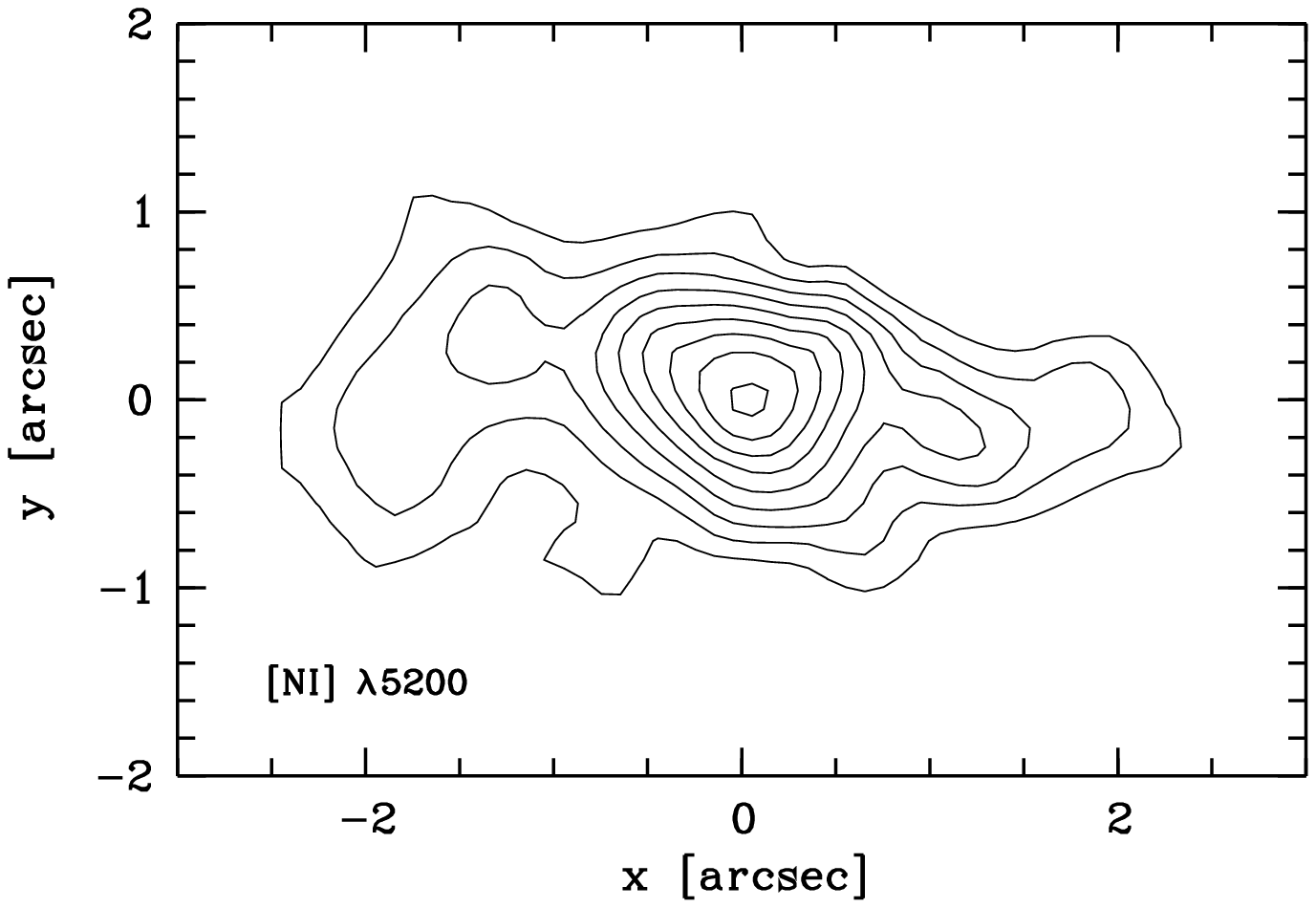,width=8.8cm}
\caption[]{\NI\lda5200 flux distribution map in NGC~2974.
The step is $3\; 10^{-16}$~\ergcmsa, and the brightest isocontour
corresponds to $27\; 10^{-16}$~\ergcmsa.}
\label{fig:NImap}
\figend

\subsubsection{Gas kinematics}

The emission-line velocity and velocity dispersion maps are shown in
Fig.~\ref{fig:gaskin}. 
The velocity field exhibits strong disturbances following the spiral-like
distribution of the gas. There is a strong hint for gas streaming on 
the inner side of the south-western arm, where the emission lines exhibit a complex
structure probably resulting from the superposition of several kinematical
components (Fig.~\ref{fig:NIIwithspec}). This widening of the lines 
($\sigma_{\rm gas} > 250$~\kms) is also observed on the inner
side of the north-eastern arm, as emphasized in the dispersion map (Fig.~\ref{fig:gaskin}).

\lfigstart{figure=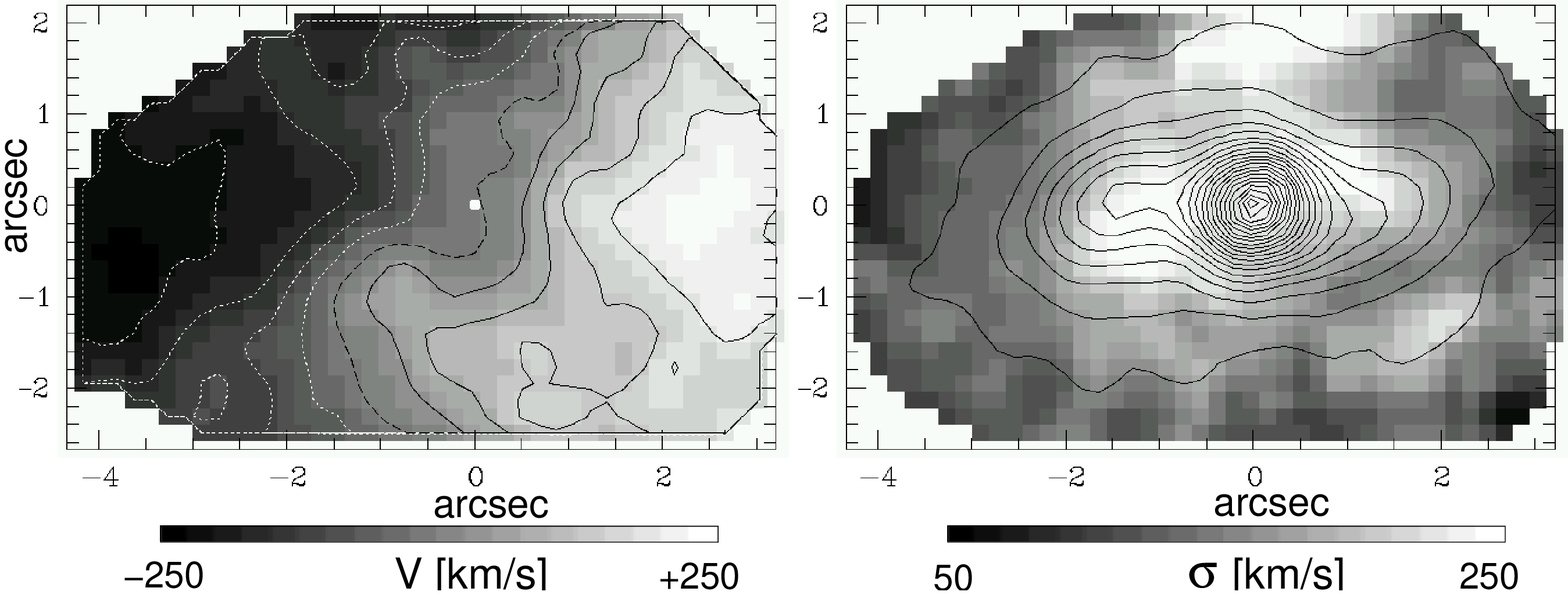,width=17.6cm}
\caption[]{Gas velocity (left) and velocity dispersion (right) maps 
for the emission lines in the \red\ datacube.
The isovelocity contours have a step of 50 \kms. The isocontours 
superimposed on the dispersion map
correspond to the \NII\lda6583 flux map (step of 10$^{-15}$~\ergcmsa).}
\label{fig:gaskin}
\lfigend

\subsection{Datacube deconvolution}
\label{sec:deconv}
Using the high-resolution {\it HST}/WFPC2 narrow-band \Ha+\NII\ image of NGC~2974
shown in Fig.~\ref{fig:HSTimages}, we applied the method
pioneered by Ferruit \etal (1999) to deconvolve our merged \tiger\
datacube (see also Emsellem \& Ferruit 2000).
Two methods were tested: a pure Lucy deconvolution, and a ``weakly guided'' Lucy deconvolution in which
the narrow-band {\it HST}/WFPC2 image is used to constrain the integrated
flux in the \red\ datacube. In both cases, we performed 300 iterations
after which the gain in resolution was found not to compensate the increase
in noise level. Differences between the results of the two deconvolutions
were not significant, except for a slight reduction of the high-frequency
features in the weakly guided case. We therefore favour the latter,
the analysis of which will be presented here.

The final resolution of the deconvolved datacube, evaluated from the
central peak (unresolved in the WFPC2 \NII$+$\Ha\ image), is $0\farcs35$
FWHM. Maps of the deconvolved datacube, namely the \NII\ and \Ha\ flux
distribution, the gas velocity and dispersion maps, are shown in
Fig.~\ref{fig:decmaps}. 

The two spiral arms are nicely revealed, as well
as the central concentration which is now highly contrasted.
The velocity field shows strong streaming motions
along the arms, with peak velocities of $-$211 and 264 \kms. 
More interestingly, the dispersion is very high on the inner side 
of both arms, confirming the picture seen before deconvolution (see
Fig.~\ref{fig:gaskin}), reaching values greater than 300 \kms. 
The gas kinematics will be further discussed in Sect.~\ref{sec:wave} in
view of a density wave model.

\lfigstart{figure=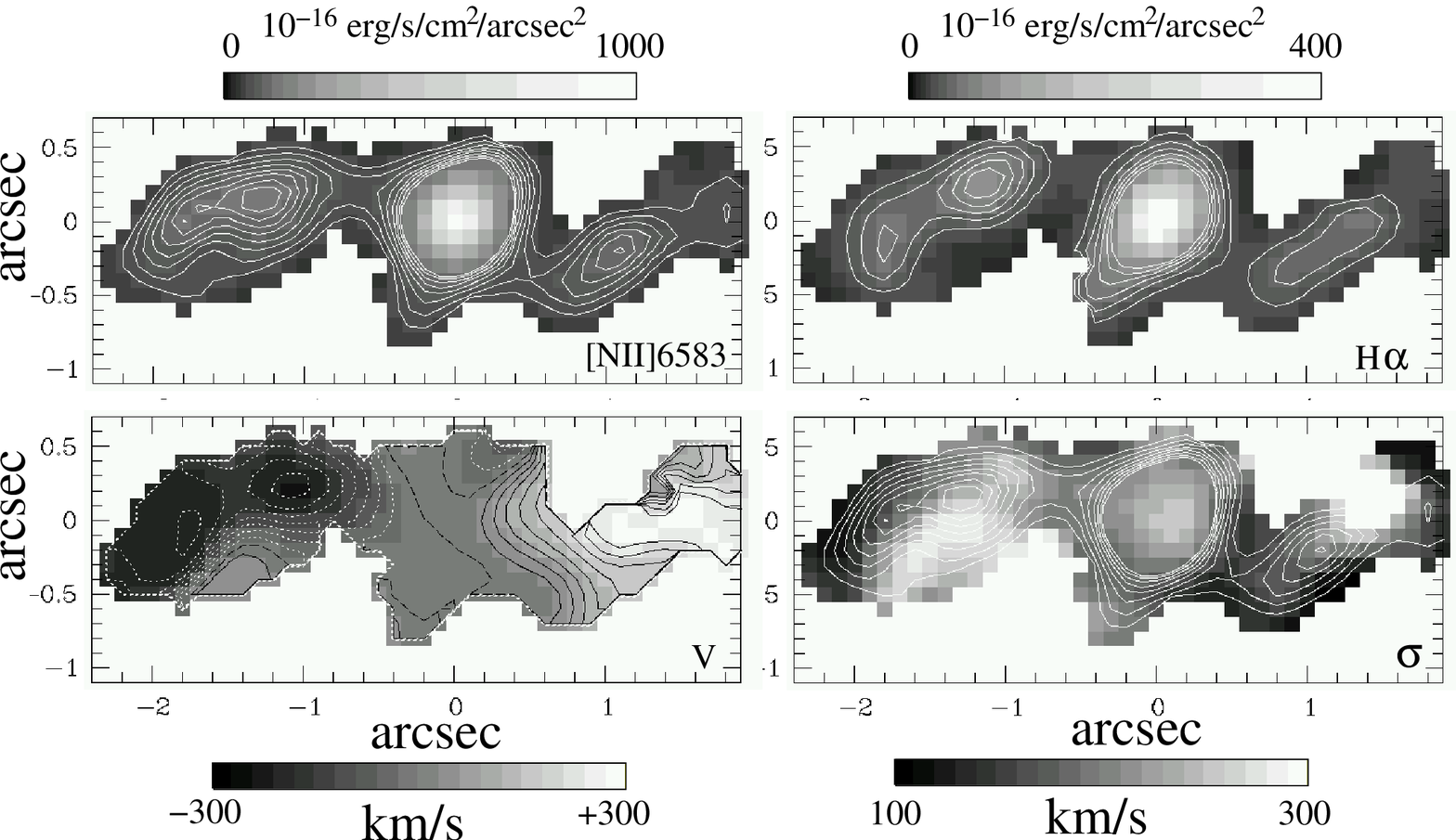,width=17.6cm}
\caption[]{Maps built from the deconvolved \tiger\ datacube:
\NII\lda6583 and \Ha\ flux distribution (top left and right panels
respectively),  gas velocity (bottom left) and velocity dispersion (bottom
right) maps. The step is 10$^{-15}$~\ergcmsa, and the lower
level contours correspond to 60 and 30 $^{-16}$~\ergcmsa for
the \NII\ and \Ha\ lines respectively.
The isovelocity contours have a step of 30 \kms. The isocontours 
superimposed on the dispersion map correspond to the \NII\lda6583 flux map 
(step of 10$^{-15}$~\ergcmsa).}
\label{fig:decmaps}
\lfigend

\section{Dynamical models}
\label{sec:model}

In this Section, we present new dynamical models for the stellar and
gas components in the central region of NGC~2974, using our
new two-dimensional data as well as published kinematics.

\subsection{The MGE luminosity model}
\label{sec:MGE}
The first step in this modelling process is to obtain a realistic
three-dimensional representation of the luminosity and mass distribution
of the galaxy from the very centre to the outer region.
We used the Multi-Gaussian Expansion method (Monnet, Bacon \& Emsellem 1992;
Emsellem \etal 1994) to build a photometric model\footnote{See Cappellari
2002 for an efficient MGE package developed within the IDL environment.}
for the deconvolved surface brightness of NGC~2974. the combination of a
ground-based $I$ band image taken at the 1.0-m JKT (data taken from Goudfrooij
\etal 1994a) and the {\it HST}/WFPC2 F814W image provided the required
wide field of view and high spatial resolution, respectively. The result of the
fit, taking into account the respective Point Spread Functions of the input images, is shown in
Fig.~\ref{fig:mgemod}. Note that the regions affected by dust extinction were masked during
the fitting procedure, although we checked that this did not affect the result of the fit
in any significant way. The spatial luminosity density is obtained by
assuming that each individual Gaussian component is uniquely deprojected as
a three-dimensional axisymmetric Gaussian (see Table~\ref{tab:mge}).

The total luminosity of the model is $2.9\ 10^{10}$\dist$^2$~\Lsun in the $I$ band.
The most flattened gaussian component in the deprojected
model is Gaussian~\#7, with an axis ratio of 0.344, which contributes to about 19\% of
the total mass in the model.

\begin{table}
\begin{flushleft}
\begin{tabular}{rrrrrrr}
\hline
\# & $I (L_{\sun}.pc^{-2}.arcsec^{-1})$ & $\sigma (\arcsec)$ & $q$ \\
\hline \hline 
1 &    807834.22 &    0.048 &  0.782 \\ 
2 &    355349.63 &    0.105 &  0.608 \\ 
3 &    62166.935 &    0.313 &  0.539 \\ 
4 &    16377.007 &    0.605 &  0.649 \\ 
5 &    3738.1343 &    1.326 &  0.598 \\ 
6 &     627.7568 &    3.735 &  0.501 \\ 
7 &     104.9081 &    9.667 &  0.344 \\ 
8 &       9.3040 &   21.829 &  0.461 \\ 
9 &       0.8656 &   55.620 &  0.466 \\ 
\hline 
\end{tabular} 
\end{flushleft} 
\caption[]{Deprojected parameters for the MGE luminosity model of NGC~2974.
From left to right: component number, maximum deprojected intensity, width as
measured by the Gaussian sigma, and axis ratio. The inclination angle was fixed
to 60\degr\ (see Sect.~\ref{sec:2I})}  
\label{tab:mge} 
\end{table}
\lfigstart{figure=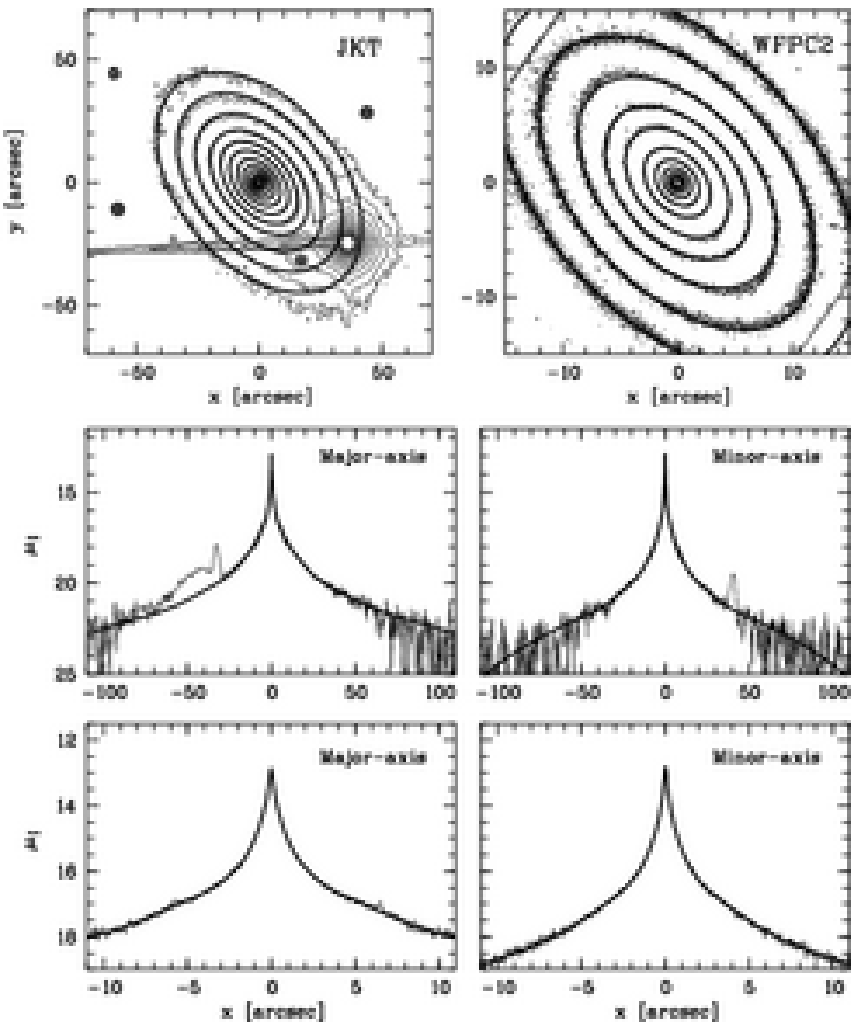,width=17.6cm}
\caption[]{Comparison between our MGE model and photometric data in the $I$
  band. The left-hand panels depict the JKT data, and the right-hand panels
  depict the WFPC2/F814W data. {\bf Top panels}: I-band isophotes (thin
  contours; step is 0.5 mag) and the corresponding MGE model isophotes 
  (thick contours). {\bf Two bottom panels}: major- and minor-axis profiles
  of the JKT and WFPC2/F814W images (thin lines) and of the MGE
  model (thick lines).}
\label{fig:mgemod}
\lfigend

\subsection{Two-integral models for the stellar component}
\label{sec:2I}
As detailed in Emsellem, Dejonghe, \& Bacon (1999, hereafter EDB99), we
constrained the inclination and mass-to-light ratio by using simple dynamical
Jeans modelling. A best-fit model was obtained with $M/L_I = 4.6\;$\dist$^{-1}$
and an inclination of $i = 60\degr$, consistent with the inclination
found by Plana \etal (1998). However, all models with values $58 \le i
\le 65$ were found to be consistent with the observed kinematics. No central
dark mass was required to fit the central velocity gradient and velocity
dispersion value at the resolution of our data (see Fig.~\ref{fig:jeansbh}). The
observed kinematics imply  
an upper limit of $2.5\; 10^8$\dist~M$_{\odot}$ to the central dark mass. This
is consistent with the mass inferred using the
M$_{\rm bh}$/$\sigma$ correlation (Ferrarese \& Merritt 2000, Gebhardt et al.\
2000). 
\figstart{figure=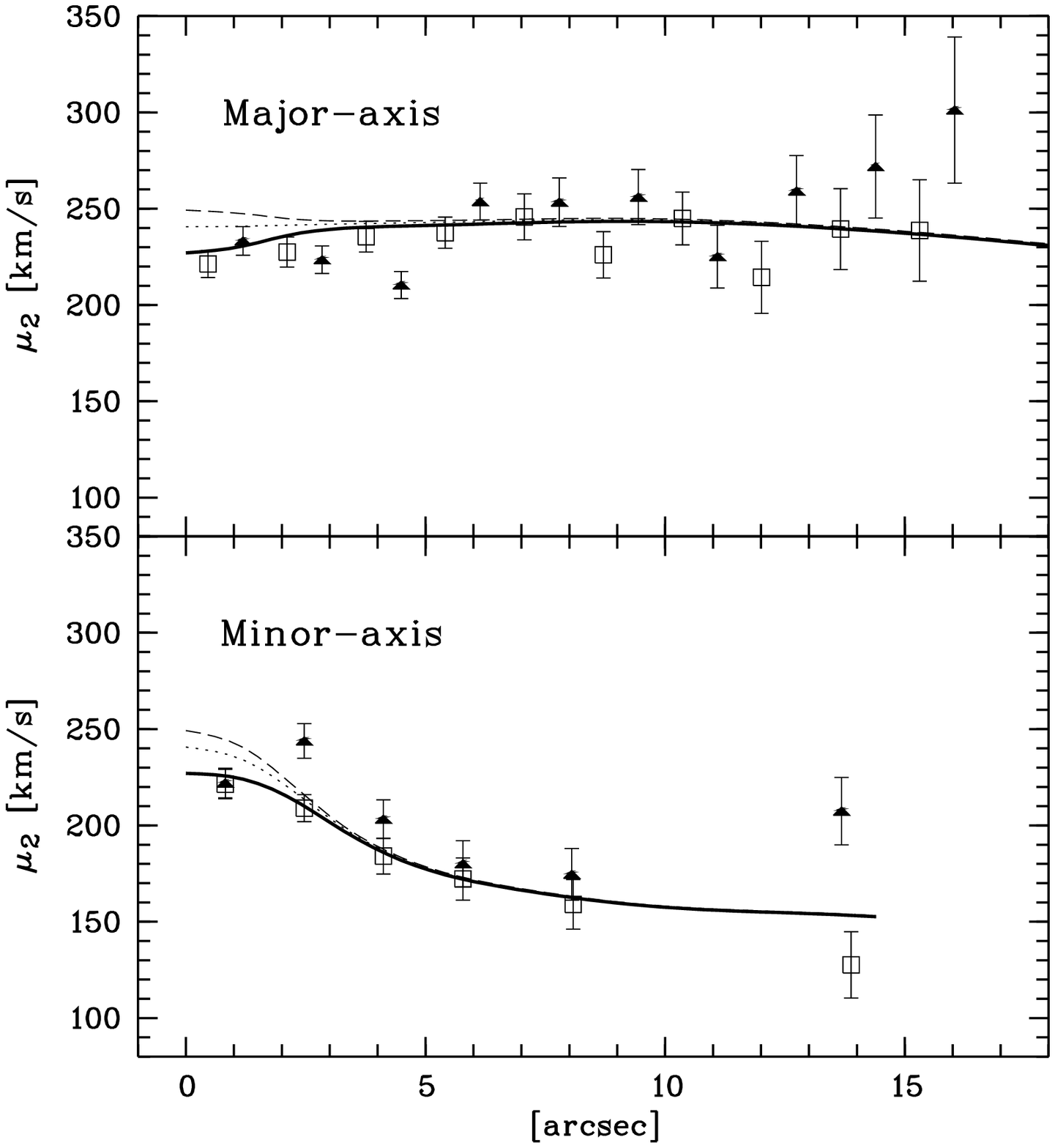,width=8.8cm}
\caption[]{Comparison between the stellar kinematics of CvdM94
  along the major (top) and minor (bottom) axis and
  isotropic Jeans models with different black hole masses (solid, dotted and
  dashed lines for 0, 3 and $5\; 10^8$~M$_{\odot}$, respectively). }
\label{fig:jeansbh}
\figend

We then derived the even part of the two-integral distribution function
$f_e(E,L_z)$ corresponding to the MGE mass model via the Hunter \& Qian
(1993) routine (details of this procedure can be found in EDB99). 
The odd part $f_o$ was parametrized with respect to the even part
using the function $h_a (\eta)$ [where $\eta \equiv J/J_{max}(E)$]: E.g., 
$f_o(E,J) = h_a(\eta) \; f_e(E,J)$ (see Eq.~1 in EDB99, and references therein).
The free parameter $a$ was taken to be the same for all components.

Line Of Sight Velocity Distributions (LOSVDs) corresponding to
the computed distribution function were then derived
on a square grid, and convolved by the appropriate kernel
to include the effect of seeing and pixel integration.
These LOSVDs were parametrized via Gauss-Hermite moments
as well as true velocity moments, and compared with 
the observed kinematics. 

\subsubsection{The best-fit model}
\label{sec:fit2I}

Our best-fit model has $M/L_I = 4.63\;$\dist$^{-1}$ with $a = 2.5$ at the centre, and $a=5$
for $E / E^{max} < 0.38$ corresponding to a circular orbital radius of $R_c >
5\arcsec$. The agreement with the CvdM94 long-slit data is excellent; a
comparison is shown in Fig.~\ref{fig:compvdm}. Although the model fits the \tiger\
maps relatively well, it can obviously not reproduce the slight tilt of the zero velocity
curve (Fig.~\ref{fig:comptig}). Our value for $M/L_I$ is larger than the one found
by CvdM94 (who found $M/L_R \sim 5.19$ or $M/L_I \sim 3.92$, normalized to
our assumed distance). However the total mass of the galaxy,
as measured on the MGE model, is $1.34\ 10^{11}$\dist$^2$~\Msun,
consistent with the value derived by CvdM94 (again for the present assumed
distance of 21.5~Mpc). We also find an excellent agreement between our value
for the predicted circular velocity  at 2\arcmin, 219~\kms, and the one derived
by CvdM94,  217~\kms. This means that  this mass to light ratio discrepancy
comes mostly from a difference of zero points in the photometry of about 0.2
magnitude. Our estimate is however significantly higher than the one obtained 
by Pizzella \etal (1997) for a triaxial geometry.

Fig.~\ref{fig:df2I} provides a representation of the total best-fit distribution function
(even plus odd parts) versus the two integrals of motions. This plot can
be compared to the one derived for NGC~3115 (EDB99).
In contrast with NGC~2974, NGC~3115
posseses rather bright and very thin disc structures, which can be easily identified
in the even part 
of the distribution function as strong peaks near the expected location for the
circular orbits ($J / J^{max} = 1$).
\figstart{figure=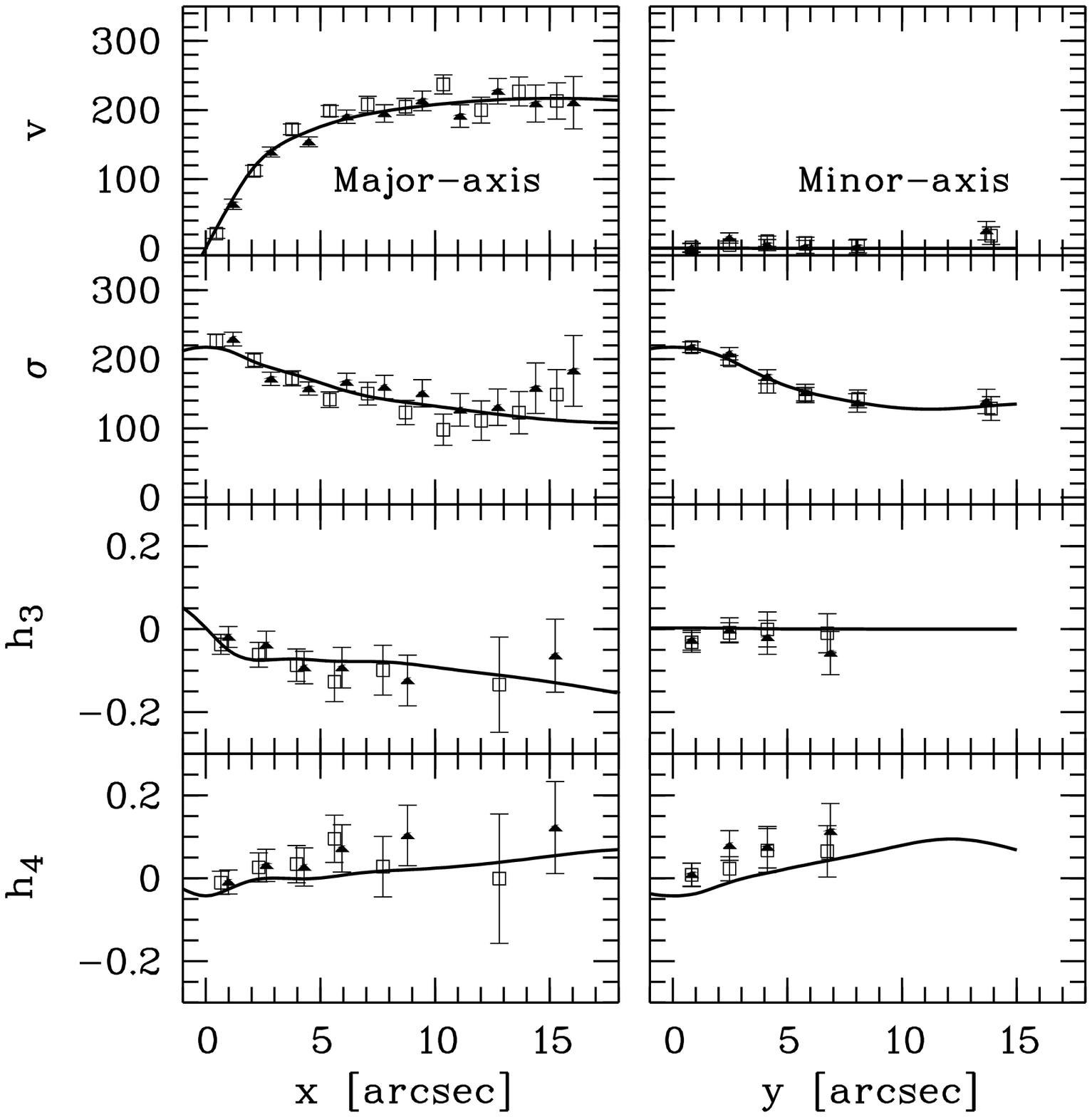,width=\hssize}
\caption[]{Comparison between our $f(E,L_z)$ model (solid lines)
and the stellar kinematics published by CvdM94 (same symbols
as in CvdM94).}
\label{fig:compvdm}
\figend
\figstart{figure=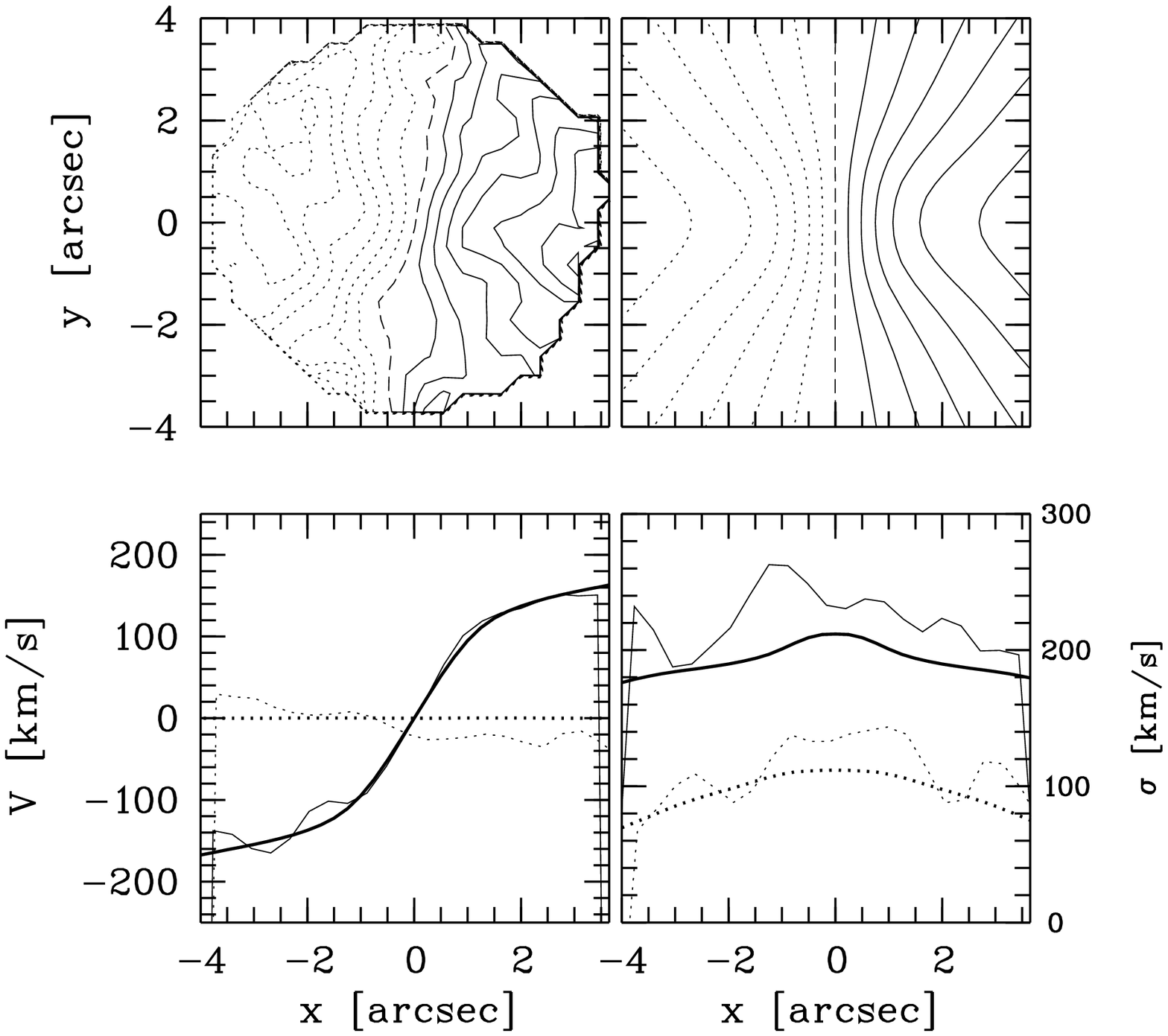,width=\hssize}
\caption[]{Comparison between our $f(E,L_z)$ model 
and the \tiger\ stellar kinematics. Top: observed (left) and modeled
velocity fields. 
Bottom: major (solid lines) and minor (dashed lines) axis velocity
(left) and dispersion (right) profiles. The thick lines correspond
to the model, the thin lines to the observed kinematics. The minor-axis
dispersion profile has been shifted for clarity. Note the slight
minor-axis velocity gradient, which is not reproduced by the axisymmetric
model.} 
\label{fig:comptig}
\figend
\figstart{figure=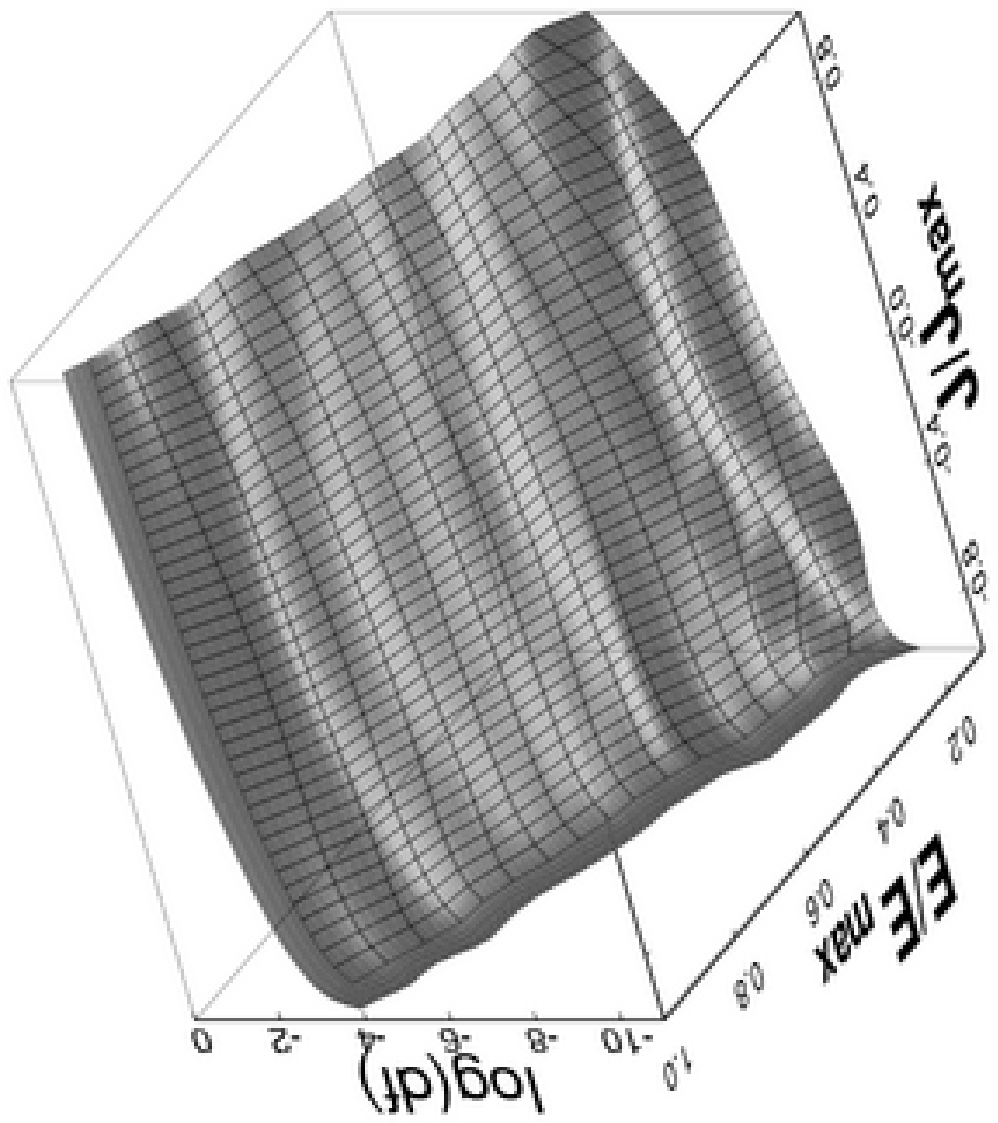,angle=-90,width=8.8cm}
\caption[]{$f(E,L_z)$ model represented as an illuminated surface of
        $\log{(df)}$ versus the normalized energy $E / E^{max}$
                and angular momentum $J / J^{max}$.}
\label{fig:df2I}
\figend

\subsection{Gaseous Density wave modelling}
\label{sec:wave}
As mentioned in Sect.~\ref{sec:gaspec}, the observed two-dimensional
morphology and velocity field of the ionized gas in the central few arcsec
show the presence of a highly contrasted, two-arm spiral structure
with strong streaming motions. At a radius of 2\arcsec, corresponding to about 220~pc at the
distance of NGC~2974, and for a velocity of about 200~\kms, the dynamical timescale is only about
$10^6$~years. The contrast and large opening of the spiral arms are therefore suggesting
that a quasi-stationary process is at play here. In what follows, we thus present our attempt to fit
the gas morphology and kinematics with a density wave model. 

\subsubsection{The geometry of the spiral arms}
\figstart{figure=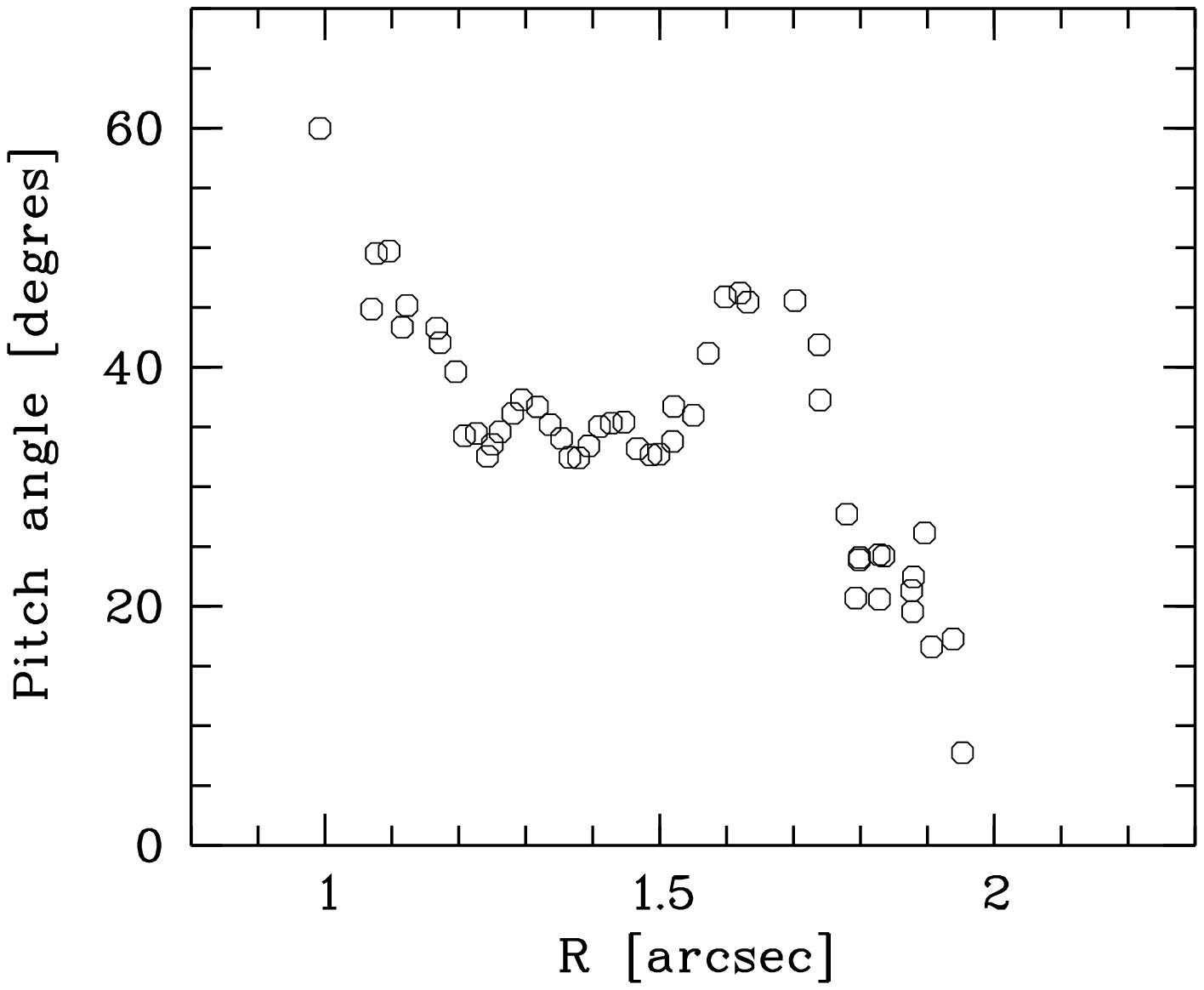,width=\hssize}
\caption[]{Pitch angle of the ionized gas spiral between a radius of
0\farcs8 and 2\arcsec, as measured from the deprojected \Ha+\NII\ WFPC2 image.}
\label{fig:pitch}
\figend

We first need to constrain the parameters of the spiral
arms, namely its geometry and amplitude. This was done
with the help of the \Ha$+$\NII\ WFPC2 image
corrected for extinction (Goudfrooij \& Emsellem 2003, in preparation). The
narrow-band image was deprojected 
assuming an inclination angle $i=60\degr$ as obtained from the stellar
dynamical model (Sect.~\ref{sec:2I}). We then estimated
the pitch angle as a function of radius. The gaseous arm on the NE side
of the centre is better defined and less patchy, probably due to the lower
extinction (the near side of the galaxy being the SE side). 
We therefore decided to preferentially use the NE arm to measure the pitch angle.
The average pitch angle is around 35\degr\ (see Fig.~\ref{fig:pitch}), with
values between 20\degr\ and 50\degr. It seems to increase towards the centre,
although the spiral is not easily traceable close to the nucleus where
the unresolved central source of emission dominates (see Fig.~\ref{fig:bar}).
We can follow the spiral up to a deprojected radius of about 0\farcs8
(i.e. apparent radius of 0\farcs4, as the connection of the spiral
with the central source unfortunately occurs close to the minor-axis
of the galaxy). Starting at a radius of 1\farcs5, an abrupt change 
in the pitch angle occurs: it decreases from 40\degr\ to almost 0\degr\
in 0\farcs5.

\subsubsection{Hints for the presence of a tumbling potential}

The total mass of ionized gas within the central 3\arcsec is
$\sim 6.8\; 10^{4}\;$\dist$^2$~M$_{\odot}$,
to be compared with a stellar mass of $0.8\; 10^{10}\;$\dist$^2$~M$_{\odot}$
derived from the MGE 
model (Sect.~\ref{sec:2I}). We can safely say that the gas
component is therefore not self-gravitating in this region.
In the limit of tightly wound spirals,
we could use the dispersion relation for non self-gravitating $m=2$ 
spirals to derive the pitch angle (assuming a value for the sound
speed $v_s$, see e.g. Englmaier \& Shlosman 2000). Such density waves are
supported inside their own Inner Lindblad Resonance
(ILR hereafter), or outside the Outer Lindblad Resonance 
(Maciejewski \etal 2002). However, the large pitch angle observed
in the case of the inner gaseous spiral arms of NGC~2974 prevents us
to use such an approximation. 

More importantly, we can also exclude the role of self-gravity as 
the source for the observed $m=2$ mode.
We therefore need to search for another driver to explain the 
highly contrasted set of arms in the inner 300~pc of NGC~2974.
The large pitch angle of the gaseous arms observed in the 
inner 2\arcsec of NGC~2974 is reminiscent
of the straight/curved gaseous and dust lanes seen in large-scale bars (e.g.
Athanassoula 1992). Such structures correspond to bar-driven shocks inside the
Corotation Resonance of the bar (CR hereafter), the dissipative component 
being driven towards the Inner Lindblad Resonance (ILR hereafter) along the 
leading edge of the bar. {\em This obviously requires the presence of an Inner 
Lindblad Resonance} (Athanassoula 1992, Maciejewski \etal 2002), 
therefore setting some constraints on the central mass profile. 

Maciejewski \etal (2002) emphasized the characteristic features of the gas flow in
a single bar. One of the main characteristics is the abrupt change
of velocity when the gas crosses the principal shocks mentioned above, flowing
from the trailing to the leading side of the arms.
This could qualitatively explain the strong streaming seen in Fig.~\ref{fig:gaskin}
and the large width of the emission line on the trailing side of the
spiral arms in NGC~2974. Maciejewski \etal (2002) also observe
gas being funelled towards the 4/1 resonance (or Ultra Harmonic
Resonance, hereafter UHR) along convergence regions.
This can create gas concentrations at the end of the principal shocks,
and in the case of NGC~2974 could account for the abrupt change
in the pitch angle at $R\sim 2\arcsec$. 
This hypothesis of the presence of an inner bar in NGC~2974 is examined in the
next section, using the formalism described in Emsellem \etal (2001).

\subsection{The bar model}
\label{sec:bar}
In Section \ref{sec:2I}, we obtained a mass model assuming axisymmetry 
which reproduces the observed stellar kinematics of NGC~2974 in
the central 18\arcsec. We can obtain a first and rough estimate of the pattern
speed $\Omega_p$ of a presumed tumbling bar by deriving an ``azimuthally
averaged'' resonance diagram for NGC~2974 using this mass model, which is shown
in Fig.~\ref{fig:resonance}. The first striking thing is that the $\Omega_p -
\kappa / 2$ radial profile predicted by this model (which does not include any
central dark mass) inevitably implies the presence of a single Inner Lindblad
Resonance. As emphasized above, the ILR is a prerequisite for the building of
shock lanes by a tumbling bar. The principal shocks should lie inside the
radius of the Ultra Harmonic 4:1 Resonance, therefore constraining $\Omega_p
\lta 1000\;$\dist$^{-1}$~\kmskpc. We also assume that the spiral arms are at
least partly outside the ILR, thus implying that  $\Omega_p \gta
500\;$\dist$^{-1}$~\kmskpc. 

We then add a bar-like component in the model by
perturbing the gravitational potential with a simple second harmonic function
of the form $\Phi(r,\theta) = -\Phi_2(r) \cos 2\theta$, where $r$ and $\theta$
are the radius and azimuth in the frame co-rotating with the bar perturbation, 
respectively. 
For $\Phi_2(r)$ ($m=2$), we use the functional form provided by the $k=3$
bar provided by Kalnajs (1976; see also Emsellem et al. 2001). 
Finally, gas orbits were calculated using the epicycle approximation for a
potential perturbed by a bar-like component (Lindblad \& Lindblad 1994;  Wada
1994). We included a `damping' term as in Emsellem \etal (2001). The free
parameters of this model are: 
the damping coefficient $\lambda$, the pattern speed of the bar $\Omega_p$,
the amplitude $q_{\rm bar}$ (in km$^{2}$ s$^{-2}$) and radius $r_{\rm bar}$
(in arcsecond) of the perturbation (see a similar but more complete approach in
Wada \& Koda 2001). For each set of parameters, we need to adjust the angle
$\theta_{bar}$, in the equatorial plane of the galaxy, between the bar and the
line of nodes.  

\figstart{figure=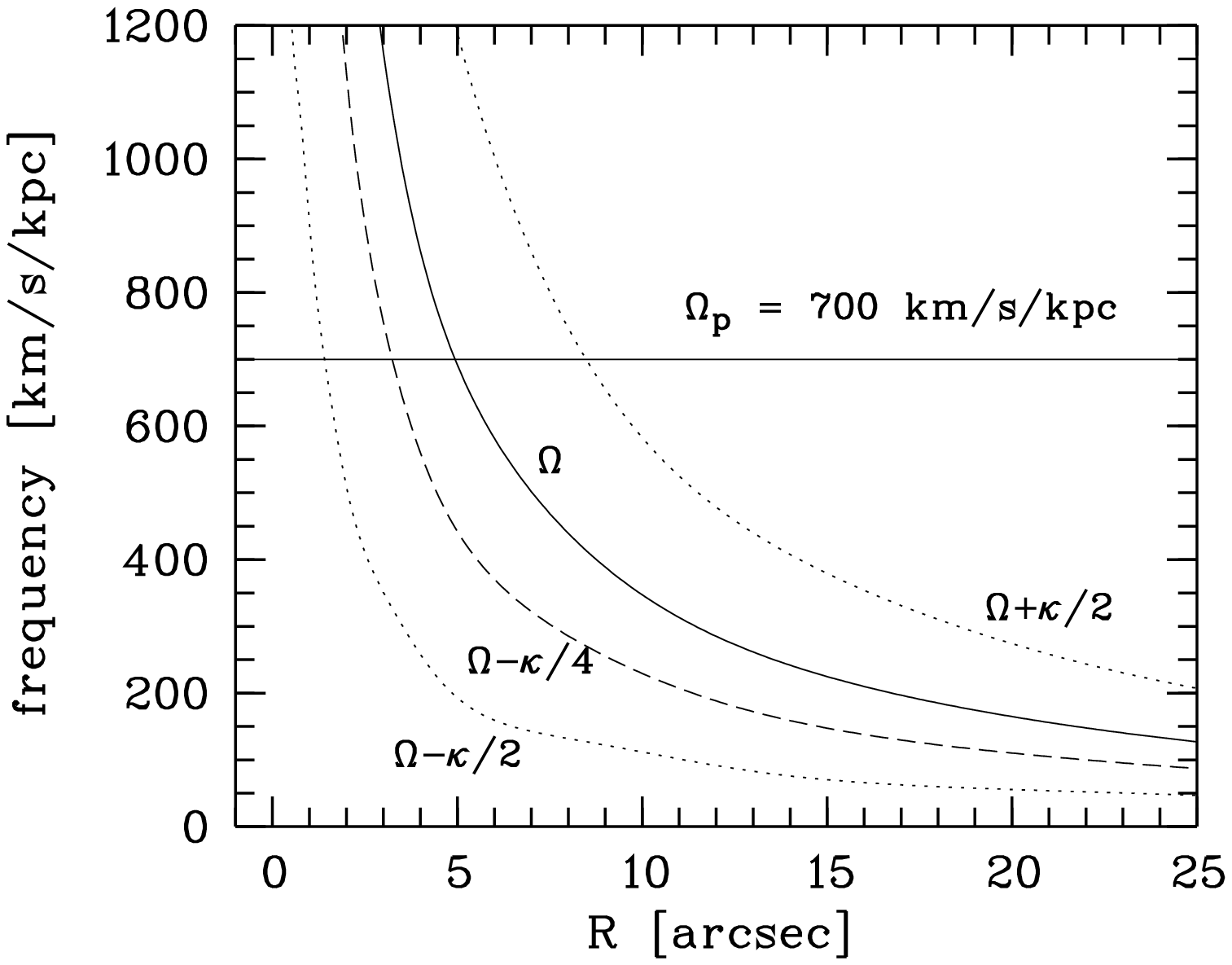,width=\hssize}
\caption[]{Resonance diagram for NGC~2974, derived from its MGE model.
Profiles of $\Omega$, $\Omega - \kappa / 2$, $\Omega + \kappa / 2$, $\Omega -
\kappa / 4$, are provided. The assumed value for the pattern speed of the bar
is shown as an horizonthal line with $\Omega_p = 700$~km/s/kpc.}
\label{fig:resonance}
\figend

\lfigstart{figure=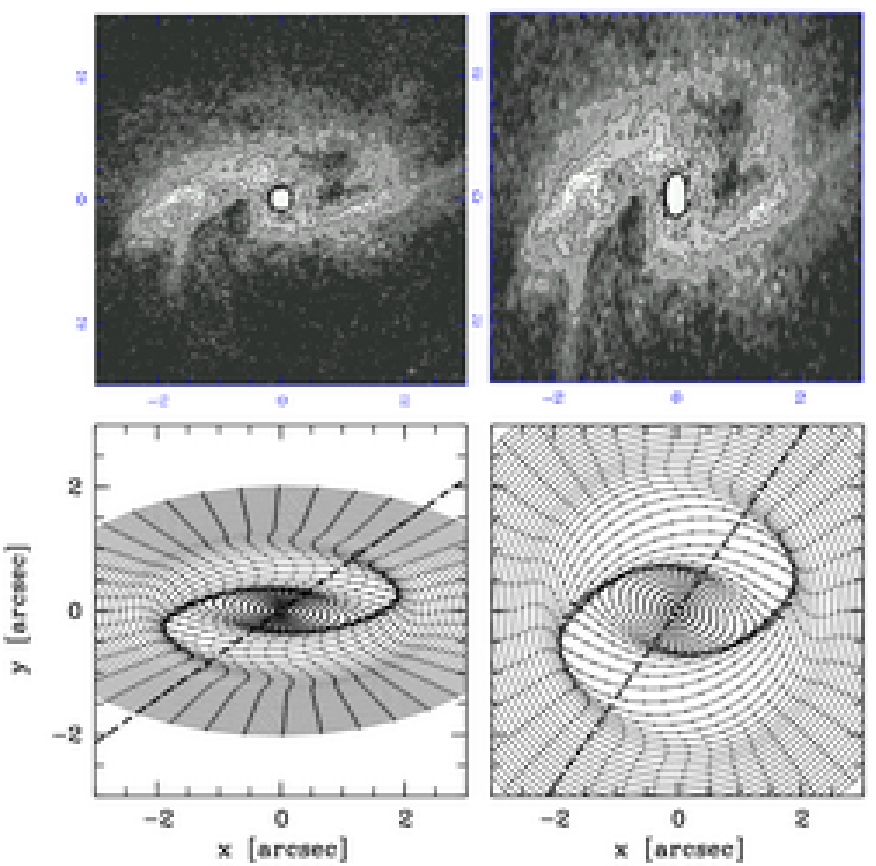,width=17.6cm}
\caption[]{Gas orbits in a model including a bar perturbation
in NGC~2974 (bottom panels) as compared to the observed WFPC2 \Ha+\NII\ image
(top panels). The left-hand panels are representations on the sky plane, the
right-hand panels correspond to deprojected views (assuming an inclination
angle of 60\degr).  The crowding of the streamlines reveal a two-arm 
spiral resembling the observed gaseous structure. The line of nodes is horizontal
and the major-axis of the bar is $\theta_{bar} = 55\degr$ away from it
(indicated by the dashed line in the lower two panels).}
\label{fig:bar}
\lfigend

\subsubsection{Comparison with the observed morphology and kinematics}

We examined systematically models with values of $\Omega_p$ within the
constrained range ($500 < \Omega_p < 1000$ in units of \kmskpc; see above) with
steps of 10~\kmskpc, for different values of $\lambda$, $q_{\rm bar}$, and
$r_{\rm bar}$.  Only values of $r_{\rm bar}$ between 2\farcs4 and 2\farcs7 can 
reproduce the change of direction of the arm at 2\farcs1; the best fit
value is $r_{\rm bar}$ = 2\farcs6 (or about 270~pc at 21.5~Mpc). 
For values of $q_{\rm bar}$ greater than $2 \times 10^5$, the pitch angle of
the corresponding arms become too high with respect to the observed values. As
the pattern speed increases, the inner side of the spiral arms moves towards 
the centre. As to $\Omega_p$, we can find reasonably good qualitative fits
to the observed structures if $500 < \Omega_p < 850$ (in \kmskpc).    

The gas orbits for a model with $\Omega_p = 700\;$\dist$^{-1}$~\kmskpc,
$q_{\rm bar} = 1 \times 10^5$ and $\theta_{bar} = 
55$\degr\ are shown in Fig.~\ref{fig:bar}, along with the observed
morphology of the ionized gas. For this model, the ILR, UHR, CR and OLR are at
radii of 1\farcs4, 3\farcs2, 4\farcs9, and 8\farcs5 respectively.
The projected major-axis of the bar is then at about 35\degr\ from the
major-axis of the galaxy (for $i = 60\degr$).
The crowding of the streamlines correspond to compression regions:
the model thus predicts the presence of shocks 
on the inner side of the arms due to the convergence of streamlines. 
This is consistent with the high velocity dispersion observed in the \tiger\ maps
(Fig.~\ref{fig:decmaps}). Note that this model corresponds to a perturbation
in the potential of less than 2\% and a maximum torque of 10\%: 
this definitely corresponds to a weak bar (see Fig.~1 of Block et al. 2002).

A detailed comparison between the observed kinematics and the one 
predicted by this model is however beyond the scope of the present study, 
as it would first require a reasonable fit of the observed line emission
distribution (all observed maps being luminosity weighted). 
This density wave model is only intended to provide a first
qualitative attempt at reproducing the observed gas distribution
and kinematics. The existing data are not sufficient to really constrain 
the input parameters of the static model we developed: we can only 
provide rough estimates of the main parameters such as: the pattern speed
$\Omega_p = 700 \pm 100$~\kmskpc, the CR radius $R_{CR} = 4.9^{+1.4}_{-0.8}$,
the angle between the bar and the line of nodes $\theta_{bar} = 55\degr \pm
20\degr$. 
It may be possible to narrow the range of possible values for
the pattern speed of the wave, as its value determines the location of the
resonances, but this would require a full hydrodynamical simulation.
A deeper high-resolution \Ha+\NII\ image would also help to follow the gaseous
distribution to greater galactocentric distances.

\section{Discussion}
\label{sec:disc}

\subsection{Dynamical models for NGC~2974}

CvdM94 assumed the presence of a hidden thin stellar disc in NGC~2974,
contributing a significant 7\% to the total visible mass of NGC~2974. In the
present study however, we solely relied on the observed surface brightness (in
the $I$ band) to build our photometric model, and did not include any
additional disc component. Our dynamical modeling shows that there is no need
for the presence of such a hidden component (see Table~\ref{tab:mge}) to fit the observed stellar
kinematics. We should however emphasize that the
detailed characteristics of the individual gaussian components provided by the
MGE formalism are (also) somewhat model-dependent, and should be taken with
some level of caution. We derived the corresponding two-integral distribution
function (strictly  speaking, {\em only\/} the even part is constrained by our
mass model) thus providing a self-consistent model. 

We note that Jeans models do not guarantee that the distribution function is
positive everywhere. They are generally mainly used to constrain the first
two  velocity moments. They should therefore be taken with caution, and
should only be used, in our view, as a useful tool to provide first
approximations on global parameters (e.g. mass-to-light ratio,
inclination). Self-consistent models should be favoured to derive LOSVDs, as
assuming the shape (e.g., Gaussian) of the contribution of individual
components (bulge, disc) may lead to misleading results. 

In our modeling of the stellar kinematics, we made the assumptions of
axisymmetry and a constant mass-to-light ratio, and neglected the effect
of dust extinction. The central gaseous kinematics (see
Sect.~\ref{sec:bar}), and the possible stellar rotation along the minor-axis
seem however to invalidate the hypothesis of axisymmetry.  We also know from
the {\it HST}/WFPC2 images that dust is indeed present in the central kpc of
NGC~2974, although its distribution is localized. The fact that a simple two-integral model fits rather well the
observed kinematics might just then be due to a lack of two-dimensional coverage
in the outer part, and to the lack of spatial resolution in the central
few arcseconds. New integral field data acquired with a larger field of view (e.g.
{\tt SAURON} on the 4.2-m WHT), better spatial resolution and signal-to-noise ratio 
should thus permit one to further constrain the stellar dynamical structure of
NGC~2974.  

\subsection{The central dark mass}

If we now interpret the large line width of the central peak of the
ionized gas ($\sigma = 460$~\kms) in terms of a rotating Keplerian disc, it
implies\footnote{This depends on the -- unknown -- radial distribution
of the gas within the central 0\farcs1.}  an enclosed mass of about $5\;10^8$~\Msun.
This is to be compared with the visible mass of $4.7 \;
10^7\;$\dist~M$_{\odot}$ from the MGE model within this radius, 
which would thus be inconsistent with the upper limit for a central dark mass 
of $2.5\; 10^8$\dist~M$_{\odot}$ (Sect.~\ref{sec:2I}). This was however
derived under the assumption of a two-integral dynamical model, which is
not valid if a tumbling bar is present. Also we would definitely need data at very
high spatial resolution to probe the central peak, and check our hypothesis of a Keplerian rotating disk.
The only instrument currently available to help unravel the dynamical nature of
the innermost region of ionized gas in NGC~2974 is the Space Telescope Imaging
Spectrograph (STIS) aboard {\it HST}. 

\subsection{The bar in NGC~2974}

The bar model presented in Sect.~\ref{sec:bar} provides a 
good representation of the observed gaseous structure and
kinematics. Within this picture, we can roughly estimate the expected
extent of the bar by assuming it is a factor of 1.2 smaller than its
CR radius: this gives a radius of 4\farcs1 or $\sim 425$~pc.
The presence of strong shocks in the gaseous spiral arms
then implies some dissipation, with a loss of angular momentum in the case of
the observed trailing spiral in NGC~2974. At the scale of the ILR radius 
(1\farcs4 or 146~pc), the dynamical timescale is short, of the order of 
$6\;10^6$~yr (the rotation period of the suggested bar being $8.7\;10^6$~yr). We therefore
expect a significant time evolution of the observed structure within a few
$10^7$~yr. In this context, the static model based on the epicyclic
approximation presented in Sect.~\ref{sec:bar} should be used with caution,
and can only serve as a guideline for future full hydrodynamical simulations.

The gaseous response to a tumbling bar for a potential with a single ILR was
examined by Fukuda, Wada \& Habe (1998) via numerical simulations.
Their corresponding run "Bb"  indeed produces trailing spiral shocks resembling
the ones we observe in NGC~2974. Their simulations 
suggest that fueling is rather inefficient in this case:\ only gas originally
within a radius about twice the ILR falls into a central structure much
smaller than the ILR (roughly 70~pc in the case of NGC~2974).
The central peak of ionized gas, which represents a reservoir of a few
$10^{3}\;$\dist$^2$~M$_{\odot}$ (see Sect.~\ref{sec:gaspec}) within a radius
of less than 0\farcs06 (6.25~pc), could be the result of such an accretion.

We finish the discussion on a perhaps more provocative note. Considering
the relatively low contrast of the presumed bar-like perturbation 
in NGC~2974, and the dust extinction, such a structure is extremely difficult 
to detect using visible broad-band photometry, even with the help
from two-dimensional stellar kinematics. 
Our result was only possible because of the availability of a
deep, high-resolution image of emission-line gas which showed the spiral
structure (Figs.~\ref{fig:HSTimages} and \ref{fig:bar}) and the simultaneous use of
two-dimensional integral-field spectroscopy. This brings up the question of
how many such bars might be hidden in other early-type galaxies? We suggest
that these might be more common than thought previously, and may provide
the possibility to funnel gas down to a scale of a few tens of parcsecs, 
a step further towards feeding the AGNs, which are commonly found in early-type
disk galaxies.
In this context, emission line gas may certainly be a more helpful tracer than
the dust features in probing the central structure of nearby galaxies, but this
then requires both high spatial resolution and two-dimensional coverage.

\section{Conclusions}
\label{sec:conc}

In this final Section, we summarize the main results of this study. 

\begin{itemize}
\item We have presented integral field spectroscopy obtained with the \tiger\
 spectrograph which allowed us to probe the stellar and gaseous distribution
 in the central 500~pc of NGC~2974. We have applied an iterative scheme
 to disentangle the relative contributions of the gaseous emission and stellar
 absorption lines from the obtained datacubes. 
\item The stellar line indices do not reveal any significant metallicity 
  gradient in the central 500 pc.
\item The ionized gas distribution exhibits a highly contrasted two-arm
  spiral structure extending up to about 2\arcsec, as well as a central
  unresolved peak of emission, representing a total gas mass of $\sim 6.8\;
  10^{4}\;$\dist$^2$~M$_{\odot}$.  
\item The gas kinematics shows strong evidence for non-circular motions,
  with complex and broad line profiles on the inner side of the spiral arms.
\item We have detected a central, unresolved line-emitting component with
  $\sigma \sim 460$~\kms, both in \Ha, and in forbidden lines (\NII\ and 
  \SII). Assuming this corresponds to an inner Keplerian disc within the
  central unresolved peak observed in the narrow-band (\Ha+\NII) {\it
    HST}/WFPC2 image, this corresponds to an equivalent mass of about
  $5\;10^8$~\Msun. 
\item We built a dynamical model
  which provides a very good fit to the \tiger\ (and available long-slit)
  stellar kinematics without requiring the addition of an extra component such
  as a disc or a central dark mass. An upper limit of $2.5\;
  10^8$\dist~M$_{\odot}$ for the mass of a potential supermassive black hole
  is derived under the assumption of a two-integral distribution function: a
  more general dynamical model could however relax this upper limit significantly.  
\item A simple density wave ($m=2$ bar) model provides a good explanation for the
  observed gas morphology and kinematics. Our mass model then inevitably
  implies the existence of a strong Inner Lindblad Resonance, which in turn
  leads to the presence of strong trailing shocks. 
\item The best-fit bar model requires a quadrupole perturbation in the
  potential of only about 2\% (with a maximum torque of 10\%), 
  with a pattern speed of $\Omega_p = 700 \pm
  100$~km.s$^{-1}$ kpc$^{-1}$. This leads to radii for the Corotation and
  Inner Lindblad Resonance of 4\farcs9 and 1\farcs4 respectively. 
\item Considering the low contrast of the required bar-like perturbation,
  its small size (270~pc in radius), and the presence of patchy dust in NGC~2974,
  such a structure would be difficult to detect photometrically in the visible. 
  The presence of such inner bars may therefore be more common in early-type disk galaxies than
  generally thought. Still, the inner bar in NGC~2974 is of sufficient strength 
  to drive gas from a few hundreds of parsecs inwards to the central tens of parsecs.  
\end{itemize}

\paragraph*{Acknowledgments.} \ \\
EE wishes to thanks Witold Maciejewski for insightful discussions.
This paper is based on observations taken with the Canada-France-Hawaii
Telescope, operated by the National Research Council of Canada, the Centre
National de la Recherche Scientifique of France, and the University of Hawaii. 
This paper is based on observations obtained with the NASA/ESA
{\it Hubble Space Telescope}, which is operated by AURA, Inc., under
NASA contract NAS 5--26555. EE is grateful to the Space Telescope
Science Institute for a visitorship during which part of this project was
carried out. PG was affiliated with the Astrophysics Division
of the Space Science Department of the European Space Agency during part of
this project.

\label{lastpage}

\begin{thebibliography}{}
\bibitem[\protect\citename{Athanassoula }1992]{atha92} 
 Athanassoula E., 1992, MNRAS, 259, 345 
\bibitem[\protect\citename{Bacon et al.\ }1995]{bac+95} 
 Bacon R., et al., 1995, A\&AS, 113, 347
\bibitem[\protect\citename{Bender }1990]{bend90} 
 Bender R., 1990, A\&A, 229, 441
\bibitem[\protect\citename{Bender et al.\ }1994]{ben+94} 
 Bender R., Saglia R. P., Gerhard O., 1994, MNRAS, 269, 785
\bibitem[\protect\citename{Bica }1988]{bica88} 
 Bica E., 1988, A\&A, 195, 76
\bibitem[\protect\citename{Block }2002]{block02} 
 Block D. L., Bournaud F., Combes F., Puerari I., Buta R., 2002, A\&A, 394, 35
 \bibitem[\protect\citename{Cappellari }2002]{capp02} 
 Cappellari M., 2002, MNRAS, 333, 400
\bibitem[\protect\citename{Chokshi \& Turner }1992]{chotur92} 
Chokshi A., Turner E. L., 1992, \mnras, 279, 421
\bibitem[\protect\citename{Cinzano \& van der Marel }1994]{cinvdm94} 
 Cinzano P., van der Marel R. P., 1994, MNRAS, 270, 325
\bibitem[\protect\citename{Davies et al.\ }1987]{davi+87} 
 Davies R. L., Burstein D., Dressler A., et al., 
 1987, ApJS, 64, 586
\bibitem[\protect\citename{Elmegreen et al.\ }1998]{elme+98} 
Elmegreen, B. G., Elmegreen D. M., Brinks E., et al., 1998, \apj, 503, L119
\bibitem[\protect\citename{Elmegreen, Chromey, Santos\ }1998]{ECS98} 
Elmegreen D. M., Chromey, F. R., Santos, M., 1998, \aj, 116, 1221
\bibitem[\protect\citename{Emsellem \& Ferruit }2000]{emse+00} 
 Emsellem E., Ferruit P., 2000, A\&A 357, 111
\bibitem[\protect\citename{Emsellem et al.\ }1994]{emse+94} 
 Emsellem E., Monnet G., Bacon R., 1994, A\&A, 285, 723
\bibitem[\protect\citename{Emsellem et al.\ }1996]{emse+96} 
 Emsellem E., Bacon R., Monnet G., Poulain P., 1996, A\&A, 312, 777
\bibitem[\protect\citename{Emsellem et al.\ }2001]{emse+01} 
 Emsellem E., Greusard D., Combes F., et al., 2001, \aaa, 368, 52
\bibitem[\protect\citename{Erwin \& Sparke }1999]{erwspa99} 
 Erwin H. P., Sparke L. S., 1999, \apj, 521, L37 
\bibitem[\protect\citename{Erwin \& Sparke }2002]{erwspa02} 
 Erwin H. P., Sparke L. S., 2002, \aj, 124, 65
\bibitem[\protect\citename{Ferrarese \& Merritt }2000]{fermer00} 
 Ferrarese L., Merritt D., 2000, ApJ, 539, L9
\bibitem[\protect\citename{Ferruit et al.\ }1999]{ferr+99} 
 Ferruit P., Wilson A. S., Falcke H., Simpson C., P\'econtal E., Durret F.,
 1999, MNRAS, 309, 1  
\bibitem[\protect\citename{Fukuda et al.\ }1998]{fuku+98} 
Fukuda H., Wada K., Habe A., 1998, \mnras, 295, 463
\bibitem[\protect\citename{Gebhardt et al.\ }2000]{gebh+00} 
 Gebhardt K., Bender R., Bower G., et al., 2000, ApJ, 539, L13
\bibitem[\protect\citename{Goudfrooij \& Emsellem }1996]{gouems96} 
 Goudfrooij P., Emsellem E., 1996, A\&A, 306, L45
\bibitem[\protect\citename{Goudfrooij et al.\ }1994a]{goud+94a} 
 Goudfrooij P., N{\o}rgaard-Nielsen H.U., Hansen L., et al., 1994a,
  A\&AS, 104, 179 
\bibitem[\protect\citename{Goudfrooij et al.\ }1994b]{goud+94b} 
 Goudfrooij P., N{\o}rgaard-Nielsen H.U., Hansen L., et al., 1994b,
  A\&AS, 105, 341
\bibitem[\protect\citename{Kim et al.\ }1988]{kim+88} 
 Kim D.-W., Guhathakurta P., van Gorkom J. H., Jura M., Knapp
  G. R., 1988, ApJ, 330, 684 
\bibitem[\protect\citename{Krist }1992]{krist92} 
 Krist J., 1992, Tinytim v2.1 User Manual (STScI) 
\bibitem[\protect\citename{Maciejewski et al.\ }2002]{mac+02} 
 Maciejewski W., Teuben P. J., Sparke L. S., Stone J. M., 2002, MNRAS, 329, 502
\bibitem[\protect\citename{Martini \& Pogge }1999]{marpog99} 
 Martini P., Pogge R. W., 1999, \aj, 118, 2646
\bibitem[\protect\citename{Monnet et al.\ }1992]{monn+92} 
 Monnet G., Bacon R., Emsellem E., 1992, A\&A, 253, 366
\bibitem[\protect\citename{Pfenniger }1991]{pfen91} 
Pfenniger D., 1991, in Sundelius B., ed, Dynamics of disc galaxies. Chalmers
 Univ.\ Press, G{\"o}teborg, p.\ 191
\bibitem[\protect\citename{Pfenniger \& Norman }1990]{pfenor90} 
Pfenniger D., Norman C. A., 1990, \apj, 363, 391 
\bibitem[Pizzella et al.(1997)]{1997A&A...323..349P} Pizzella, A.,
Amico, P., Bertola, F., Buson, L.~M., Danziger, I.~J., Dejonghe, H., 
Sadler, E.~M., Saglia, R.~P., de Zeeuw, P.~T., Zeilinger, W.~W.,
1997, A\&A, 323, 349 
\bibitem[\protect\citename{Plana et al.\ }1998]{plan+98} 
 Plana H., Boulesteix J., Amram Ph., Carignan C., Mendes de Oliveira C., 1998,
 A\&A 128, 75  
\bibitem[\protect\citename{Regan \& Mulchaey }1999]{regmul99} 
 Regan M., Mulchaey J. S., 1999, \aj,  117, 2676
\bibitem[\protect\citename{Rieke \& Lebofsky }1985]{rieleb+85} 
 Rieke G. H., Lebofsky M. J., 1985, ApJ, 288, 618
\bibitem[\protect\citename{Rousset }1992]{rous92} 
 Rousset, A., 1992, Ph.D. Thesis, Univ. de Saint-Etienne
\bibitem[\protect\citename{Shlosman et al.\ }1989]{shlo+89} 
 Shlosman I., Frank J., Begelman M. C., 1989, \nat, 338, 45
\bibitem[Tonry et al.(2001)]{2001ApJ...546..681T} Tonry, J.~L., Dressler, 
A., Blakeslee, J.~P., Ajhar, E.~A., Fletcher, A.~B., Luppino, G.~A., 
Metzger, M.~R., \& Moore, C.~B.\ 2001, \apj, 546, 681 
\bibitem[\protect\citename{Wada }1994]{wada94} 
 Wada, K., 1994, PASJ, 46, 165
\bibitem[\protect\citename{Wada \& Koda }2001]{wadkod01} 
 Wada, K., Koda, J., 2001, PASJ, 53, 1163 
\bibitem[\protect\citename{Zeilinger et al.\ }1996]{zeil+96} 
 Zeilinger W. W., Pizzella A., Amico P., et al., 1996, A\&AS,
  120, 257
\end{thebibliography}
\end{document}